\documentclass[letterpaper]{IEEEtran}
\usepackage{amsfonts}
\usepackage[dvips]{graphicx}
\usepackage{epsfig,latexsym}
\usepackage{float}
\usepackage{indentfirst}
\usepackage{amsmath}
\usepackage{amssymb}
\usepackage{times}
\usepackage{subfigure}
\usepackage{stfloats,algorithm,algorithmic,bm}
\usepackage{psfrag}
\usepackage{cite}
\usepackage{subfigure}
\usepackage{}
\usepackage{amsfonts,subfigure,multicol,color,verbatim,
graphicx,cite,epsfig,amssymb,amsmath}
\newtheorem{theorem}{Theorem}
\newtheorem{lemma}[theorem]{Lemma}
\newtheorem{proposition}[theorem]{Proposition}

\begin{document}

\markboth{IEEE JOURNAL ON SELECTED AREAS IN COMMUNICATIONS, VOL. 33,
NO. X, Jan. 2015} {Hu et al: Network coding in social networks
\ldots}

\title{
\textsc{Network Coded Multi-Hop Wireless Communication Networks:
Channel Estimation and Training Design}
\author{\IEEEauthorblockN{Mugen~Peng,~\IEEEmembership{Senior~Member,~IEEE,},
Qiang Hu, Xinqian Xie, Zhongyuan Zhao, and H. Vincent Poor, ~\IEEEmembership{Fellow,~IEEE}}
\thanks{Manuscript received Apr. 17, 2013; revised Sep. 15, 2014; accepted
Nov. 07, 2014. The work of M. Peng, Q. Hu, X. Xie, and Z. Zhao was
supported in part by the National Natural Science Foundation of
China (Grant No. 61222103), the National High Technology Research
and Development Program of China (Grant No. 2014AA01A701), and the
Chinese State Major Science and Technology Special Project (Grant
No. 2013ZX03001001). The work of H. V. Poor was supported in part by
the U.S. National Science Foundation under Grant ECCS-1343210.}
\thanks{M. Peng, Q. Hu, X. Xie, and Z. Zhao are with the Key Laboratory of Universal
Wireless Communications, Ministry of Education, Beijing University of Posts \text{\&}
Telecommunications, Beijing, China. Mugen Peng is also with the School of
Engineering and Applied Science, Princeton University, Princeton,
NJ, USA (e-mail: {\tt pmg@bupt.edu.cn}; {\tt xxmbupt@gmail.com};
{\tt hq760001570@gmail.com}; {\tt zyzhao@bupt.edu.cn}).}
\thanks{H. V. Poor is with the School of Engineering
and Applied Science, Princeton University, Princeton, NJ, USA
(e-mail: {\tt poor@princeton.edu}). }}}

\maketitle

\begin{abstract}
User cooperation based multi-hop wireless communication networks
(MH-WCNs) as the key communication technological component of mobile
social networks (MSNs) should be exploited to enhance the capability
of accumulating data rates and extending coverage flexibly. As one
of the most promising and efficient user cooperation techniques,
network coding can increase the potential cooperation performance
gains among selfishly driven users in MSNs. To take full advantages
of network coding in MH-WCNs, a network coding transmission strategy
and its corresponding channel estimation technique are studied in
this paper. Particularly, a $4-$hop network coding transmission
strategy is presented first, followed by an extension strategy for
the arbitrary $2N-$hop scenario ($N\geq 2$). The linear minimum mean
square error (LMMSE) and maximum-likelihood (ML) channel estimation
methods are designed to improve the transmission quality in MH-WCNs.
Closed form expressions in terms of the mean squared error (MSE)
for the LMMSE channel estimation method are derived, which allows
the design of the optimal training sequence. Unlike the LMMSE
method, it is difficult to obtain closed-form MSE expressions
for the nonlinear ML channel estimation method. In order to
accomplish optimal training sequence design for the ML method, the
Cram\'{e}r-Rao lower bound (CRLB) is employed. Numerical results are
provided to corroborate the proposed analysis, and the results
demonstrate that the analysis is accurate and the proposed methods are
effective.
\end{abstract}


\IEEEpeerreviewmaketitle

\section{Introduction}

With the dramatic evolution of mobile communication systems and
the rapid rise in the use of advanced mobile devices, the original
web-based social networks have comprehensively penetrated into the
mobile platform in recent years, motivating the newly emerged research
field of mobile social networks (MSNs)\cite{1}. Currently, a large
proportion of global communication traffic is contributed by
user-generated activities associated with MSNs, e.g., instant
messages, document sharing, and interactions within friend circles.
MSNs have been characterized as pervasive and omnipotent mobile
communication platforms involving social relationships via which
users can search, share and deliver data anytime and
anywhere\cite{2}.

Since MSNs encourage new modes of socially driven information
flow and provide a backbone for modern communications, they have
motivated considerable research interest for more than a decade. Most
earlier studies of MSNs concentrated on human interactions
and relations. More recently, considerable work has been devoted to
addressing the intersection and interplay of online social networks
and wireless communications from a technological viewpoint \cite{7}
\cite{8}, and rapid development in wireless communications
 have essentially driven the expansion of MSNs \cite{9}.

Despite this research effort, there are few existing works
addressing the interplay between technological networks and social
networks in MSNs. Some promising attempts at exploiting this
interplay have been reported in \cite{10}, including security and
privacy problems\cite{11}, network design \cite{12} and
efficient resource management\cite{13}. Even so, there are still
many unexplored significant technological challenges in the
development of MSNs, which could lead to optimal design for socially
based technological networks to offer better user experiences and
services.

Stimulated by socially driven incentives, users in MSNs are
willing to interact with others via sharing or delivering
information voluntarily, which is referred to as initiative user
cooperation\cite{14}. Thereby, cooperative communication, regarded
as an effective way to accumulate data rates and extend coverage
flexibly\cite{15}, can improve the performance of MSNs by taking
advantages of socially enabled collaborative features. Due to
physical constraints, information exchange between two socially linked
users in MSNs may need the cooperation of intermediate users in
the underlying technological network, resulting in multi-hop wireless
communication networks (MH-WCNs). As one of the most promising and
efficient user cooperation techniques, network coding \cite{16} has
significant potential to further improve the performance of MH-WCNs.
The intense research effort on network coded MH-WCNs has established
that significant performance gains can be obtained with network coding. Sengupta,
\emph{et al.} analyzed throughput improvements obtained by network
coding for MH-WCNs in \cite{17}. Network coding can also be
incorporated into communication protocols for enhancing the
reliability and speed of data gathering in smart grids\cite{18} and
wireless sensor networks\cite{19}. Opportunistic network coding for
optimization of routing strategies was introduced for wireless mesh
networks in \cite{20}. Further, the feasibility of network coding
for applications in MSNs and its capacity to facilitate user
cooperation have been demonstrated in \cite{21}. However, to the
best of our knowledge, how to design efficient network coded
transmission strategies and how to implement network coding in
practical radio channels for MH-WCNs are still not straightforward,
which are key challenges to promote the commercial development of
MSNs.

Since network coding enables each node to use coding operations on
several packets\cite{22}, the nodes in network coded MH-WCNs need to
mitigate the redundant self-interference through acquiring the
necessary channel state information (CSI). Therefore, suitable
network coding strategies and corresponding channel
estimation methods are indispensable for network coded MH-WCNs.
In \cite{23}, Gao,\emph{ et al.} provided a preliminary study of
training based channel estimation issues for network coded two-way
relay networks (TWRNs). Further, channel estimation in TWRNs
with power allocation at intermediate nodes was further
investigated in \cite{24}. Afterwards, maximum a posteriori probability
(MAP) based channel estimation algorithm was developed for TWRNs in
\cite{25}. The authors in \cite{WSN_Hist1} have shown that
performance of network coded TWRNs significantly degrades under
imperfect CSI conditions. The existing works for channel
estimation and training design in network coding are mainly
concentrated on $2-$hop TWRNs, where the network coding strategy is
relatively simple as the data decoding and channel estimation
are only required at the desired nodes. Moreover, the performance of
traditional training designs is good enough for $2-$hop TWRNs owing
to no requirements of channel estimation at the intermediate node.
However, the extensions of network coding strategy, training design,
and channel estimation from $2-$hop TWRNs to MH-WCNs are not
straightforward. Actually, these extensions are challenging due to
the fact that self-interference cancelation and acquisition
of CSI in MH-WCNs are required at both intermediate and desired
nodes. Further, the general network coding transmission strategy for
the arbitrary hops in MH-WCNs is still a topic for research.
Consequently, it is of interest to develop an effective network coding
strategy, and its corresponding channel estimation and training design
for MH-WCNs.

Driven by solving these aforementioned extensions, an adaptive
network coded multi-hop strategy is presented in this paper to
improve the transmission spectral efficiency of MH-WCNs, which
results in enhancing the performance of MSNs. Further, channel
estimation and corresponding training design schemes are
proposed to improve the transmission quality of MH-WCNs and
to support practical implementations of network coded MH-WCNs. The main
contributions are listed as follows.

\begin{itemize}
\item A network coded multi-hop transmission strategy for MH-WCNs is proposed,
which can enable the periodic reception of data symbols at the
desired nodes, and achieve twofold spectral efficiency performance
gains compared with the traditional point-to-point strategy.
\item To guarantee that each intermediate node obtains the required CSI accurately,
a suitable training scheme is designed, where not only the desired
nodes but also the selected intermediate nodes are allowed to
transmit training sequences, which can reduce overall
resource consumption needed for training.
\item Focusing on the special $4-$hop scenario, both the linear minimum mean
square error (LMMSE) and maximum-likelihood (ML) channel estimation
methods for composite channel coefficients at desired nodes are
proposed. Furthermore, the optimal training sequences aiming at
minimizing the LMMSE-based mean squared error (MSE) and the ML-based
Cram\'{e}r-Rao lower bound (CRLB) performance metrics are derived
for LMMSE and ML methods, respectively. This network coded $4-$hop
transmission strategy and the corresponding channel estimation
methods are extended to the arbitrary $2N-$hop scenario ($N \geq
2$). Simulation results indicate the effectiveness of the
proposed estimators in improving the estimation accuracy and
achieving significant performance gains over the traditional
point-to-point estimation scheme.
\end{itemize}

The rest of this paper is organized as follows. In Section II, the
relationship between MSNs and MH-WCNs is explained, and the network
coding strategy in $4-$hop WCNs is presented. Section III describes
the LMMSE and ML channel estimation methods for the composite
channel coefficients. The training design techniques for the channel
estimation methods are discussed in Section IV. In Section V, the
network coding strategy and the corresponding training design
schemes in a general $2N-$hop WCNs are described. Simulation results
are presented in Section V to verify the proposed network coded
transmission strategy, and the corresponding channel estimation and
training design schemes. Finally, we offer conclusions in Section
VI, followed by the related proofs in the appendix.

\textbf{Notation:} Vectors and matrices are denoted by boldface small and
capital letters, respectively. The transpose, complex conjugate,
Hermitian, inverse, and pseudo-inverse of the matrix $\mathbf{A}$
are denoted by $\mathbf{A}^{T}$, $\mathbf{A}^{*}$, $\mathbf{A}^{H}$,
$\mathbf{A}^{-1}$ and $\mathbf{A}^{\dag}$, respectively.
$\textup{tr}(\mathbf{A})$ is the trace of $\mathbf{A}$, and
$\textup{diag}\{\mathbf{A}\}$ denotes a diagonal matrix constructed
from diagonal elements of $\mathbf{A}$. $[\mathbf{A}]_{ij}$
represents the $(i,j)$-th element of $\mathbf{A}$ and $\mathbf{I}$
is the identity matrix. $\|\mathbf{a}\|$ denotes the 2-norm of the
vector $\mathbf{a}$. $\mathfrak{R}\{\cdot\}$ and
$\mathfrak{I}\{\cdot\}$ denote the real and the imaginary part of
the complex argument; $\mathcal{E}\{\cdot\}$ denotes the statistical
expectation.

\section{System Model for 4-Hop WCNs}

An MSN consisting of both wireless and social contacts can be
characterized as a two-layer heterogeneous network, which is
depicted in Fig. 1. The communication networking layer (lower layer)
represents the overall radio communication links amongst different
users, while the upper one depicts the social communication links
amongst users. The social layer is relational to the virtual
communication as any two users can potentially form a contact link.
On the contrary, the radio communication links are constrained by
physical constraints, such as transmission distance, radio
interference and transmission power, which results in possible
difficulty in establishing a direct communication link between any
two arbitrary users. Hence, link formation between two arbitrary users in the
social layer of MSNs may transform into a multi-hop
communication link with the help of user cooperation in the
underlying communication networking layer. This fact indicates the
strong interplay between these two layers and the emerging demand
for the design of underlying multi-hop radio communication links to
enable social communication in the upper layer.

\begin{figure}[t]
\center\vspace*{1em}
\includegraphics[width=2.7in]{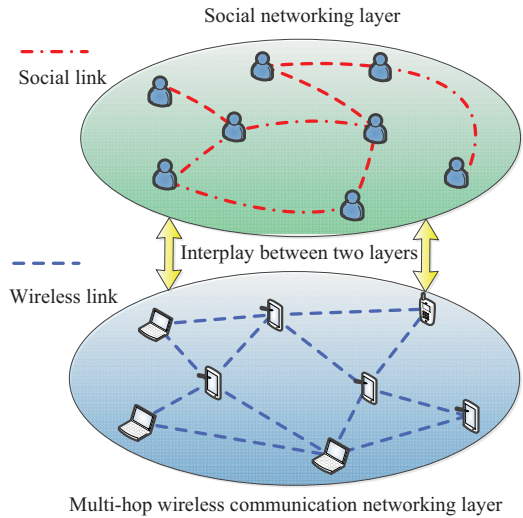}
\caption{System architecture for MSNs}
\label{fig1}\vspace*{-1em}
\end{figure}

To guarantee the quality and robustness of social links in MSNs,
radio communication links should be connected adaptively. Denoting
each user or each data source as a communication node, and regarding
the wireless connection between adjacent users as a communication
link, two arbitrary users can communicate with each other in a
bidirectional manner. Thus, the multi-hop communication links in the
underlying communication networking layer can be modeled as an
MH-WCN from the technological viewpoint. To improve the spectral
efficiency of MH-WCNs, network coding can be exploited. For the convenience of
understanding, we firstly consider simple $4-$hop MSNs as a
paradigm in this paper, where two nodes are randomly selected
to exchange information with each other via three intermediate
communication nodes.

The system model for the network coded $4-$hop WCNs, consisting of two
communication nodes $\mathbb{T}_{1}$ and $\mathbb{T}_{2}$ and three
cooperative communication nodes $\mathbb{R}_{1}$, $\mathbb{R}_{2}$
and $\mathbb{R}_{3}$, is shown in Fig. 2. Each node is equipped with a
single antenna, and the half-duplex communication protocol is
adopted in the intermediate nodes for information exchange. The
transmission powers for nodes $\mathbb{T}_{1}$, $\mathbb{T}_{2}$,
$\mathbb{R}_{1}$, $\mathbb{R}_{2}$, $\mathbb{R}_{3}$ are set to be
$P_{1}$, $P_{2}$, $P_{r1}$, $P_{r2}$, $P_{r3}$, respectively. The
data symbol sent by $\mathbb{T}_{i}$ is denoted by $x_{i}(i = 1,
2)$. The radio channel between any two adjacent communication nodes
is assumed to be quasi-flat fading, and the channel response of the
$i$-th hop from the communication node $\mathbb{T}_{1}$ to the
intermediate node $\mathbb{R}_{3}$ is denoted by $h_{i}\in\mathcal
{C}\mathcal {N}(0,\sigma^{2}_{2i-1})(i = 1, 2)$. Similarly,
$g_{i}\in\mathcal {C}\mathcal {N}(0,\sigma^{2}_{2i})(i = 1, 2)$
denotes the channel coefficient of the $i$-th hop from
$\mathbb{T}_{2}$ to $\mathbb{R}_{3}$. Since this paper is focused on
channel estimation and training design for MH-WCNs, the
communication nodes are assumed to be synchronous, which can be
fulfilled through synchronization technologies such as the global
positioning system (GPS), post-facto synchronization \cite{28}, etc.
\begin{figure}[t]
\center\vspace*{-2em}
\includegraphics[width=2.7in]{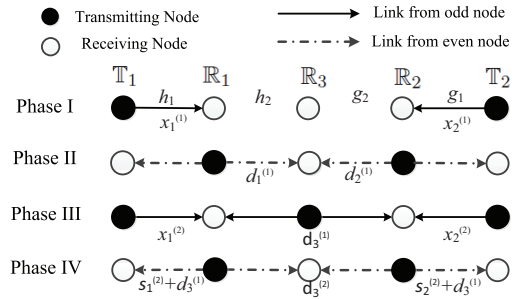}
\caption{Network coding transmission strategy for $4-$hop WCNs.}
\label{fig:system model}\vspace*{-2em}
\end{figure}

For traditional $4-$hop WCNs, if $\mathbb{T}_{1}$ and
$\mathbb{T}_{2}$ want to exchange information with each other, each
desired node occupies 4 time phases to deliver its information to
the destination node, and there are a total of 8 phases to fulfill
this information exchange. To improve the spectral efficiency, the
network coding transmission strategy is proposed in this paper,
which can complete the information exchange in 4 phases. In the
first phase, the desired nodes $\mathbb{T}_{1}$ and $\mathbb{T}_{2}$
transmit their information $x_1^{(1)}$ and $x_2^{(1)}$
simultaneously to $\mathbb{R}_{1}$ and $\mathbb{R}_{2}$,
respectively. $\mathbb{R}_{1}$ and $\mathbb{R}_{2}$ amplify their
received signals and broadcast them to neighboring nodes in the
second phase. In the third phase, $\mathbb{T}_{1}$ and
$\mathbb{T}_{2}$ transmit new information $x_1^{(2)}$ and
$x_2^{(2)}$ simultaneously to $\mathbb{R}_{1}$ and $\mathbb{R}_{2}$,
respectively, and $\mathbb{R}_{3}$ broadcasts the compound signal
of $x_1^{(1)}$ and $x_2^{(1)}$ to both $\mathbb{R}_{1}$ and
$\mathbb{R}_{2}$. Both $\mathbb{R}_{1}$ and $\mathbb{R}_{2}$
broadcast the received compound signal and new information (i.e.,
$x_1^{(2)}$ and $x_2^{(2)}$) to adjacent nodes in the fourth phase.
Based on the received information, $\mathbb{T}_{1}$ and
$\mathbb{T}_{2}$ can exchange their information with each other,
and $\mathbb{R}_{3}$ can obtain the compound signal of $x_1^{(2)}$
and $x_2^{(2)}$.

\subsection{Network Coding Transmission Strategy in $4-$Hop WCNs}

As noted above, to
accomplish the information exchange between the desired nodes
$\mathbb{T}_{1}$ and $\mathbb{T}_{2}$, data
transmission for the network coded $4-$hop WCNs is divided into $4$
phases. In the first phase, the desired nodes $\mathbb{T}_{1}$ and
$\mathbb{T}_{2}$ transmit simultaneously, and the
intermediate nodes $\mathbb{R}_{1}$ and $\mathbb{R}_{2}$ receive

\begin{equation}\label{1}
d_{1}^{(1)}=h_{1}x_1^{(1)}+n_{1}^{(1)},
\end{equation}
and
\begin{equation}\label{2}
d_{2}^{(1)}=g_{1}x_{2}^{(1)}+n_{2}^{(1)},
\end{equation}
respectively. $n_{i}^{(1)}$ represents additive white Gaussian
noise (AWGN) with zero mean and $\sigma^{2}_{n}$ variance at
node $i$, where the superscript represents the round number of data
transmission. $\mathbb{R}_{i}\left(i\!=\!1,2\right)$ amplifies
the received signal by a fixed gain $
\tilde{\alpha}_{i}\!=\!\sqrt{\frac{P_{ri}}{P_{i}\sigma_{i}^{2}\!
+\!\sigma_{n}^{2}}}$ and broadcasts
$\alpha_{i}d_{i}^{\left(1\right)}$ to its neighboring nodes in the
second phase. Then, the superimposed signal $\mathbb{R}_{3}$
received can be expressed as
\begin{equation}\label{5}
d_{3}^{(1)}=\tilde{\alpha}_{1}h_{1}h_{2}d_{1}^{(1)}+\tilde{\alpha}_{2}g_{1}g_{2}d_{2}^{(1)}
+\tilde{\alpha}_{1}h_{1}n_{1}^{(1)}+\tilde{\alpha}_{2}g_{2}n_{2}^{(1)}+n_{3}^{(1)}.
\end{equation}

In order to accomplish the information exchange between the desired
nodes $\mathbb{T}_{1}$ and $\mathbb{T}_{2}$, the intermediate node
$\mathbb{R}_{3}$ needs to send back the compound signal
$d_{3}^{(1)}$ to the desired nodes, which simply reverses the
forgoing data transmission process. Firstly, $\mathbb{R}_{3}$
broadcasts the scaled network coded signal $\alpha_{3}d_{3}^{(1)}$
to $\mathbb{R}_{1}$ and $\mathbb{R}_{2}$. At the same time,
$\mathbb{R}_{1}$ and $\mathbb{R}_{2}$ also receive the new
information $x_{1}^{(2)}$ and $x_{2}^{(2)}$ transmitted from the two
desired nodes $\mathbb{T}_{1}$ and $\mathbb{T}_{2}$ for the second
data transmission round, respectively. Then, the compound signals
received at $\mathbb{R}_{1}$ and $\mathbb{R}_{2}$ in the third phase
are

\begin{equation}\label{6}
d_{1}^{(2)}=\overbrace{h_{1}x_{1}^{(2)}+n_{1}^{(2)}}^{s_{1}^{(2)}}+\alpha_{3}h_{2}d_{3}^{(1)},
\end{equation}
\begin{equation}\label{7}
d_{2}^{(2)}=\overbrace{g_{1}x_{2}^{(2)}+n_{2}^{(2)}}^{s_{2}^{(2)}}+\alpha_{3}g_{2}d_{3}^{(1)},
\end{equation}
respectively, where the amplified scaling factor can be expressed as
$\alpha_{3}=\sqrt{\frac{P_{r3}}{P_{r1}\sigma_{3}^{2}+P_{r2}\sigma_{4}^{2}+\sigma_{n}^{2}}}$.
The received signals at $\mathbb{R}_{1}$ or $\mathbb{R}_{2}$ consist
of the new transmitted signal $x_{i}^{(2)}$ and the superimposed
network coded signal $d_{3}^{(1)}$.

Next, in order to complete the first data exchange round
between $\mathbb{T}_{1}$ and $\mathbb{T}_{2}$ and continue the
second data transmission round simultaneously, $\mathbb{R}_{1}$ and
$\mathbb{R}_{2}$ need to broadcast the amplified compound signals
$d_{1}^{(2)}$ and $d_{2}^{(2)}$ to $\mathbb{T}_{1}$,
$\mathbb{T}_{2}$ and $\mathbb{R}_{3}$. As a consequence, in the
fourth phase, $\mathbb{T}_{1}$ and $\mathbb{T}_{2}$ receive
\begin{equation}\label{8}
y_{1}=\alpha_{1}h_{1}s_{1}^{(2)}+\alpha_{1}\alpha_{3}h_{1}h_{2}d_{3}^{(1)}+n_{y_{1}},
\end{equation}
\begin{equation}\label{9}
y_{2}=\alpha_{2}g_{1}s_{2}^{(2)}+\alpha_{2}\alpha_{3}g_{1}g_{2}d_{3}^{(1)}+n_{y_{2}},
\end{equation}
respectively. The amplified scaling factors can be represented as
$\alpha_{1}=\sqrt{\frac{P_{r1}}{P_{1}\sigma_{1}^{2}+P_{r3}\sigma_{3}^{2}+\sigma_{n}^{2}}}$
and
$\alpha_{2}=\sqrt{\frac{P_{r2}}{P_{2}\sigma_{2}^{2}+P_{r3}\sigma_{4}^{2}+\sigma_{n}^{2}}}$.
As $d_{3}^{(1)}$ consists of the data symbols $x_{1}^{(1)}$ and
$x_{2}^{(1)}$ transmitted by the desired nodes in the first round,
the desired nodes can extract the data sent from others by canceling
the data sent by themselves, e.g., $\mathbb{T}_{1}$ can obtain
$x_{2}^{(1)}$ sent from $\mathbb{T}_{2}$ by deleting $x_{1}^{(1)}$
and $x_{1}^{(2)}$ transmitted by itself in the first round and
second round, respectively. Meanwhile, the compound signals
received at $\mathbb{R}_{3}$ are combinations of the new transmitted
signal $s_{3}^{(2)}$ and the redundant superimposed network coded
signal $d_{3}^{(1)}$ back from the previous round, which can be written
as
\begin{equation}\label{10}
d_{3}^{(2)}=\overbrace{\alpha_{1}h_{2}s_{1}^{(2)}+\alpha_{2}g_{2}s_{2}^{(2)}
+n_{3}^{(2)}}^{s_{3}^{(2)}}+\left(
\alpha_{1}h_{2}^{2}+\alpha_{2}g_{2}^{2}\right)d_{3}^{(1)}.
\end{equation}

It is inappropriate for $\mathbb{R}_{3}$ to continue broadcasting the
redundant signal $d_{3}^{(1)}$, which results in energy
dissipation due to the worthless information transmission and
overlapping of the newly transmitting signal $s_{3}^{(2)}$.
Consequently, the redundant signal $d_{3}^{(1)}$ is required to be
removed leaving only $s_{3}^{(2)}$, so that the network coded
data can continue to be sent out. The proposed network coded
transmission strategy can achieve twofold spectral efficiency
performance gains over the traditional point-to-point strategy. In
order to remove the redundant signal in $\mathbb{R}_{3}$, CSI
estimates $\hat{h}_{2}$ and $\hat{g}_{2}$
corresponding to ${h}_{2}$ and ${g}_{2}$ are needed, which can be
acquired by standard radio channel estimation techniques. For this purpose,
it is necessary to design the training process for $4-$hop
WCNs.

\subsection{Training Design in $4-$Hop WCNs}

Considering the self-interference cancelation required at the
intermediate node $\mathbb{R}_{3}$, a completed form of the training
design scheme is neeeded for both desired nodes and the intermediate
node $\mathbb{R}_{3}$ to transmit the pilot training sequences. The
training scheme should avoid the excessive consumption of resources
incurred in the point-to-point scheme, where the training sequences
are required for every hop. Denoting the $L$ symbol length training
sequences from $\mathbb{T}_{1}$, $\mathbb{T}_{2}$ and
$\mathbb{R}_{3}$ as $\mathbf{t}_{1}$, $\mathbf{t}_{2}$ and
$\mathbf{t}_{r}$, respectively, we further define
$Q_{i}\!=\!\|\mathbf{t}_{i}\|^{2} (i = 1, 2)$ and
$Q_{r}\!=\!\|\mathbf{t}_{r}\|^{2}$ for notational simplicity.
Following the four time phases for the proposed network coding
transmission strategy, each training round is divided into $4$
phases as shown in Fig. 3.
\begin{figure}[t]
\center\vspace*{-1em}
\includegraphics[width=2.7in]{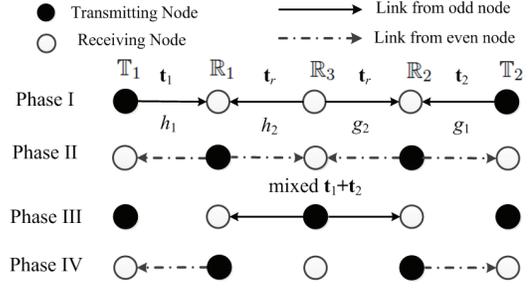}
\caption{Training design in $4-$hop WCNs.}
\label{fig3}\vspace*{-1em}
\end{figure}

In the first phase, the two desired nodes and $\mathbb{R}_{3}$
transmit simultaneously, and the received signals at
$\mathbb{R}_{1}$ and $\mathbb{R}_{2}$ are expressed as
\begin{equation}\label{11}
\mathbf{r}_{1}=h_{1}\mathbf{t}_{1}+h_{2}\mathbf{t}_{r}+\mathbf{n}_{1}^{(1)},
\end{equation}
\begin{equation}\label{12}
\mathbf{r}_{2}=g_{1}\mathbf{t}_{2}+g_{2}\mathbf{t}_{r}+\mathbf{n}_{2}^{(1)},
\end{equation}
respectively, where $\mathbf{n}_{i}^{(1)}$ is the corresponding
$L\times 1$ dimensional AWGN vector. Then, $\mathbb{R}_{i}$
amplifies $\mathbf{r}_{i} (i = 1, 2)$ by the fixed gain $\alpha_{i}$
and forwards $\alpha_{i}\mathbf{r}_{i}$ to its neighboring nodes in
the second phase. The received training signals at $\mathbb{T}_{1}$,
$\mathbb{T}_{2}$ and $\mathbb{R}_{3}$ can be written as

\begin{align}\label{13W}
\mathbf{z}_{1}&=\alpha_{1}\mathbf{T}_{1}\mathbf{h}_{1}^{(1)}+\alpha_{1}h_{1}
\mathbf{n}_{1}^{(1)}+\mathbf{n}_{\mathbf{z}_{1}},
\end{align}
\begin{align}\label{14}
\mathbf{z}_{2}&=\alpha_{2}\mathbf{T}_{2}\mathbf{g}_{1}^{(1)}+\alpha_{2}g_{1}
\mathbf{n}_{1}^{(1)}+\mathbf{n}_{\mathbf{z}_{2}},
\end{align}

\begin{align}\label{15}
\mathbf{r}_{3}\!=\!\alpha_{1}h_{1}h_{2}\mathbf{t}_{1}\!+\!\alpha_{2}g_{1}g_{2}\mathbf{t}_{2}\!
+\!\left(\alpha_{1}h_{2}^{2}\!+\!\alpha_{2}g_{2}^{2}\right)\mathbf{t}_{r}\!+\!\tilde{\mathbf{n}}_{3}^{(1)},
\end{align}
respectively, where $\mathbf{T}_{1}\!=\![\mathbf{t}_{1},
\mathbf{t}_{r}]$, $\mathbf{T}_{2}\!=\![\mathbf{t}_{2},
\mathbf{t}_{r}]$, $\mathbf{h}_{1}^{(1)}\!=\![h_{1}^{2},
h_{1}h_{2}]^{T}$, $\mathbf{g}_{1}^{(1)}\!=\![g_{1}^{2},
g_{1}g_{2}]^{T}$, and
$\tilde{\mathbf{n}}_{3}^{(1)}=\alpha_{1}h_{2}\mathbf{n}_{1}^{(1)}+\alpha_{2}g_{2}
\mathbf{n}_{1}^{(1)}+\mathbf{n}_{3}^{(1)}$.
After the first training transmission round, $\mathbb{T}_{1}$ and
$\mathbb{T}_{2}$ perform channel estimation to acquire
$\hat{\mathbf{h}}_{1}^{(1)}$ and $\hat{\mathbf{g}}_{1}^{(1)}$,
respectively. Meanwhile, $\mathbb{R}_{3}$ needs to obtain an
estimate of $\alpha_{1}h_{2}^{2}+\alpha_{2}g_{2}^{2}$ for
self-interference cancelation. Implementing the least squares (LS)
estimation method for
$\mathbf{h}_{r1}=[h_{1}h_{2},g_{1}g_{2},\alpha_{1}h_{2}^{2}+\alpha_{2}g_{2}^{2}]^{T}$,
the estimate can be  written as
\begin{equation}\label{17}
\hat{\mathbf{h}}_{r1}=\mathbf{\Lambda}_{0}^{-1}\mathbf{T}_{r}^{\dagger}\mathbf{r}_{3}
=\mathbf{h}_{r1}+\mathbf{\Lambda_{0}}^{-1}\mathbf{T}_{r}^{\dagger}\tilde{\mathbf{n}}_{3}^{(1)},
\end{equation}
where
$\mathbf{T}_{r}=[\mathbf{t}_{1},\mathbf{t}_{2},\mathbf{t}_{r}]$,
$\mathbf{T}_{r}^{\dagger}=\left(\mathbf{T}_{r}^{H}\mathbf{T}_{r}\right)^{-1}\mathbf{T}_{r}^{H}$,
and
$\mathbf{\Lambda_{0}}=\textup{diag}\left\{\alpha_{1},\alpha_{2},1\right\}$.

Since $\mathbf{t}_{r}$ is useless for the channel estimation at the
desired nodes, the training signals transmitted in the third phase
only retain $\mathbf{t}_{1}$ and $\mathbf{t}_{2}$. The corresponding
channel parameters can be written as
$\mathbf{h}_{r}=[h_{1}h_{2},g_{1}g_{2}]$. In this case,
$\mathbb{R}_{3}$ broadcasts the re-organized training signals based
on LS estimation, which can be expressed as
\begin{equation}\label{18}
\tilde{\mathbf{r}}_{3}=\tilde{\alpha}_{3}\mathbf{T\Lambda}\hat{\mathbf{h}}_{r}
=\tilde{\alpha}_{3}\left(\alpha_{1}h_{1}h_{2}\mathbf{t}_{1}+\alpha_{2}g_{1}g_{2}
\mathbf{t}_{2}+\tilde{\mathbf{n}}_{r}\right),
\end{equation}
where $\mathbf{T}=[\mathbf{t}_{1},\mathbf{t}_{2}]$, and
$\mathbf{\Lambda}=\textup{diag}\{\alpha_{1},\alpha_{2}\}$. $\tilde{\mathbf{n}}_{r}
=\mathbf{T}\left(\mathbf{T}^{H}\mathbf{T}
\right)^{-1}\mathbf{T}^{H}\tilde{\mathbf{n}}_{3}^{(1)}$
is the LS estimation error, which can be regarded as the residual
noise.

The total noise power can be calculated as
\begin{equation}\label{19}
\mathcal{E}\left\{\tilde{\mathbf{n}}_{r}^{H}\tilde{\mathbf{n}}_{r}\right\}=2\left(\alpha_{1}^{2}
\sigma_{3}^{2}+\alpha_{2}^{2}\sigma_{4}^{2}+1\right)\sigma_{n}^{2}.
\end{equation}

The parameter $\tilde{\alpha}_{3}$ can be explicitly written as
\begin{equation}\label{20}
\tilde{\alpha}_{3}=\sqrt{\frac{LP_{r}}{\alpha_{1}\sigma_{1}^{2}\sigma_{3}^{2}LP_{1}
+\alpha_{2}\sigma_{2}^{2}\sigma_{4}^{2}LP_{2}+2\left(\alpha_{1}^{2}\sigma_{3}^{2}
+\alpha_{2}^{2}\sigma_{4}^{2}+1\right)\sigma_{n}^{2}}}.
\end{equation}

Finally, the received signal at $\mathbb{T}_{1}$ can be expressed as
\begin{equation}\label{21}
\mathbf{z}_{3}=\alpha_{1}h_{1}h_{2}\tilde{\mathbf{r}}_{3}+\tilde{\mathbf{n}}_{\mathbf{z}_{3}},
\end{equation}
where
$\tilde{\mathbf{n}}_{\mathbf{z}_{3}}=\alpha_{1}h_{1}\mathbf{n}_{1}^{(2)}+\mathbf{n}_{\mathbf{z}_{3}}$
is the $L\times 1$ dimensional equivalent noise vector.

By substituting \eqref{18} into \eqref{21}, $\mathbf{z}_{3}$ can be
rewritten as
\begin{equation}\label{22}
\mathbf{z}_{3}=\alpha_{1}\tilde{\alpha}_{3}\mathbf{T}\mathbf{\Lambda}\boldsymbol{\theta}
+\tilde{\mathbf{n}},
\end{equation}
where $\boldsymbol{\theta}=[h_{1}^{2}h_{2}^{2},
h_{1}h_{2}g_{1}g_{2}]^{T}$ and
$\tilde{\mathbf{n}}=\alpha_{1}\tilde{\alpha_{3}}h_{1}h_{2}\tilde{\mathbf{n}}_{r}^{(1)}
+\alpha_{1}h_{1}\mathbf{n}_{1}^{(2)}+\mathbf{n}_{\mathbf{z}_{3}}$.

\section{Channel Estimation for $4-$Hop WCNs}

In this section, both LMMSE and ML estimation schemes
are analyzed to obtain radio
channel coefficients at each node in $4-$hop WCNs. The
correlation coefficient between the training sequences
$\mathbf{t}_{1}$ and $\mathbf{t}_{2}$ sent by
$\mathbb{T}_{1}$ and $\mathbb{T}_{2}$, respectively, can be written as
$\rho=\mathbf{t}_{1}^{H}\mathbf{t}_{2}/\sqrt{Q_{1}Q_{2}}$.

\subsection{LMMSE Channel Estimation Scheme}

The second order statistics $\sigma_{i}^{2}\;(i=1,\dots,4)$ and
$\sigma_{n}^{2}$ are assumed to be known by the transmitters. Following
the standard approach in \cite{26}, according to the received
training signal in \eqref{22}, the radio channel
$\boldsymbol{\theta}$ can be estimated as
$\hat{\boldsymbol{\theta}}$ via LMMSE estimation, and
$\hat{\boldsymbol{\theta}}$ can be directly derived as

\begin{equation}\label{24}
\hat{\boldsymbol{\theta}}=\mathcal{E}\left\{\boldsymbol{\theta}\mathbf{z}_{3}^{H}
\right\}\big(\mathcal{E}\left\{\mathbf{z}_{3}\mathbf{z}_{3}^{H}\right\}\big)^{-1}\mathbf{z}_{3}.
\end{equation}

Substituting \eqref{22} into \eqref{24}, the radio channel
estimate can be re-expressed as
\begin{equation}\label{(25)}
\hat{\boldsymbol{\theta}}=\alpha_{1}\tilde{\alpha}_{3}\mathbf{R}_{\boldsymbol{\theta}}
\mathbf{\Lambda}\mathbf{T}^{H}\left(\alpha_{1}^{2}\tilde{\alpha}_{3}^{2}\mathbf{T\Lambda}\mathbf{R}_{\boldsymbol{\theta}}
\mathbf{\Lambda}\mathbf{T}^{H}+\mathbf{R}_{\tilde{\mathbf{n}}}\right)^{-1}\mathbf{z}_{3},
\end{equation}
where
$\mathbf{R}_{\boldsymbol{\theta}}\!=\!\mathcal{E}\{\boldsymbol{\theta}\boldsymbol{\theta}^{H}\}\!
=\!\textup{diag}\{\underbrace{4\sigma_{1}^{4}\sigma_{3}^{4}}_{\sigma_{\theta_{1}}^{2}},
\underbrace{\sigma_{1}^{2}\sigma_{2}^{2}\sigma_{3}^{2}\sigma_{4}^{2}}_{\sigma_{\theta_{2}}^{2}}\}$.

The corresponding MSE is given by
\begin{equation}\label{(26)}
\sigma_{\boldsymbol{\theta}}^{2}=\mathrm{tr}\big\{\left(\mathbf{R}_{\boldsymbol{\theta}}^{-1}
+\alpha_{1}^{2}\tilde{\alpha}_{3}^{2}\mathbf{\Lambda}\mathbf{T}^{H}\mathbf{R}_{\tilde{\mathbf{n}}}^{-1}
\mathbf{T}\mathbf{\Lambda}\right)^{-1}\big\},
\end{equation}
where the covariance matrix of the equivalent noise is
$\mathbf{R}_{\tilde{\mathbf{n}}} =
\sigma_{n}^{2}\big(\xi\mathbf{I}_{N}+\alpha_{1}^{2}\tilde{\alpha_{3}}^{2}\sigma_{1}^{2}\sigma_{3}^{2}
\varepsilon\mathbf{T}\left(\mathbf{T}^{H}\mathbf{T}\right)^{-1}\mathbf{T}^{H}\big)$.
To simplify notation, we can define
$\varepsilon=\left(2\alpha_{1}^{2}\sigma_{3}^{2}+\alpha_{2}^{2}\sigma_{4}^{2}+1\right)$,
and $\xi=\left(1+\alpha_{1}^{2}\sigma_{1}^{2}\right)$. By
substituting the equivalent noise covariance matrix
$\mathbf{R}_{\tilde{\mathbf{n}}}$ into
$\mathbf{R}_{\mathbf{z}_{3}}$, the explicit expression of
$\mathbf{R}_{\mathbf{z}_{3}}$ can be written in
\begin{equation}\label{(27)}
\mathbf{R}_{\mathbf{z}_{3}}=\sigma_{n}^{2}\xi\left(\mathbf{I}_{N}+A_{1}\mathbf{t}_{1}
\mathbf{t}_{1}^{H}+A_{2}\mathbf{t}_{2}\mathbf{t}_{2}^{H}+A_{3}\mathbf{t}_{1}\mathbf{t}_{2}^{H}
+A_{3}^{*}\mathbf{t}_{2}\mathbf{t}_{1}^{H}\right),
\end{equation}
where $
A_{1}=\frac{\alpha_{1}^{2}\tilde{\alpha}_{3}^{2}}{\xi}\left(\frac{\alpha_{1}\sigma_{\theta_{1}}^{2}}
{\sigma_{n}^{2}}+\frac{\sigma_{1}^{2}\sigma_{3}^{2}\varepsilon}{\left(1-|\rho|^{2}\right)Q_{1}}\right)$,
$A_{2}=\frac{\alpha_{1}^{2}\tilde{\alpha}_{3}^{2}}{\xi}\left(\frac{\alpha_{2}\sigma_{\theta_{2}}^{2}}
{\sigma_{n}^{2}}+\frac{\sigma_{1}^{2}\sigma_{3}^{2}\varepsilon}{\left(1-|\rho|^{2}\right)Q_{2}}\right)$,
and
$A_{3}=-\alpha_{1}^{2}\tilde{\alpha}_{3}^{2}\frac{\rho\sigma_{1}^{2}\sigma_{3}^{2}\varepsilon}
{\left(1-|\rho|^{2}\right)\sqrt{Q_{1}Q_{2}}\xi}$.

By expanding \eqref{(25)}, $\hat{\theta_{1}}$ and $\hat{\theta_{2}}$
can be estimated respectively as
\begin{align}\label{(28)}
\hat{\theta_{1}}=\alpha_{1}^{2}\tilde{\alpha}_{3}\sigma_{\theta_{1}}^{2}\mathbf{t}_{1}^{H}
\mathbf{R}_{\mathbf{z}_{3}}^{-1}\mathbf{z}_{3},
\end{align}
\begin{align}\label{(28-2)}
\hat{\theta_{2}}=\alpha_{1}\alpha_{2}\tilde{\alpha_{3}}\sigma_{\theta_{2}}^{2}
\mathbf{t}_{2}^{H}\mathbf{R}_{\mathbf{z}_{3}}^{-1}\mathbf{z}_{3}.
\end{align}

By substituting \eqref{(27)} into \eqref{(28)} and \eqref{(28-2)},
the corresponding channel estimates can be rewritten as
\begin{align}\label{(30)}
\hat{\theta_{1}}&=\frac{\alpha_{1}^{2}\tilde{\alpha_{3}}\sigma_{\theta_{1}}^{2}}
{\tau\xi\sigma_{n}^{2}}\big[(1+A_{2}Q_{2}+A_{3}^{*}\rho\sqrt{Q_{1}Q_{2}})\mathbf{t}_{1}^{H}\nonumber\\
&-(A_{3}^{*}Q_{1}+A_{2}\rho^{*}\sqrt{Q_{1}Q_{2}}\mathbf{t}_{2}^{H})\big]\mathbf{z}_{3},
\end{align}
\begin{align}\label{(31)}
\hat{\theta_{2}}&=\frac{\alpha_{1}\alpha_{2}\tilde{\alpha_{3}}\sigma_{\theta_{2}}^{2}}
{\tau\xi\sigma_{n}^{2}}\big[(1+A_{1}Q_{1}+A_{3}^{*}\rho\sqrt{Q_{1}Q_{2}})\mathbf{t}_{2}^{H}\nonumber\\
&-(A_{3}Q_{2}+A_{1}\rho\sqrt{Q_{1}Q_{2}}\mathbf{t}_{2}^{H})\big]\mathbf{z}_{3},
\end{align}
where
$\tau\!=\!1\!+\!A_{1}Q_{1}\!+\!A_{2}Q_{2}\!+\!2A_{3}\rho^{*}\sqrt{Q_{1}Q_{2}}\!
+\!\left(A_{1}A_{2}\!-\!|A_{3}|^{2}\right)xQ_{1}Q_{2}$.

The MSEs of $\theta_{1}$ and $\theta_{2}$ are defined as
$e_{\theta_{1}}$ and $e_{\theta_{2}}$, respectively, as shown in
\begin{align}\label{(32)}
e_{\theta_{1}}=\mathcal{E}\left\{|\Delta_{a}|^{2}\right\}&=\sigma_{\theta_{1}}^{2}
-\alpha_{1}^{4}\tilde{\alpha_{3}}^{2}\sigma_{\theta_{1}}^{4}\mathbf{t}_{1}^{H}\mathbf{R}_
{\mathbf{z}_{3}}^{-1}\mathbf{t}_{1},\\
e_{\theta_{2}}=\mathcal{E}\left\{|\Delta_{b}|^{2}\right\}&=\sigma_{\theta_{2}}^{2}
-\alpha_{1}^{2}\alpha_{2}^{2}\tilde{\alpha_{3}}^{2}\sigma_{\theta_{2}}^{4}\mathbf{t}_{2}^{H}
\mathbf{R}_{\mathbf{z}_{3}}^{-1}\mathbf{t}_{2}.
\end{align}

Similarly, the quantities
$\mathbf{t}_{1}^{H}\mathbf{R}_{\mathbf{z}_{3}}^{-1}\mathbf{t}_{1}$
and
$\mathbf{t}_{2}^{H}\mathbf{R}_{\mathbf{z}_{3}}^{-1}\mathbf{t}_{2}$
can be written as
\begin{equation}\label{(33)}
\mathbf{t}_{1}^{H}\mathbf{R}_{\mathbf{z}_{3}}^{-1}\mathbf{t}_{1}=\frac{Q_{1}+A_{2}
\left(1-|\rho|^{2}\right)Q_{1}Q_{2}}{\tau\xi\sigma_{n}^{2}},
\end{equation}
\begin{equation}\label{(34)}
\mathbf{t}_{2}^{H}\mathbf{R}_{\mathbf{z}_{3}}^{-1}\mathbf{t}_{2}=\frac{Q_{2}+A_{1}
\left(1-|\rho|^{2}\right)Q_{1}Q_{2}}{\tau\xi\sigma_{n}^{2}}.
\end{equation}

\subsection{ML Estimation Scheme}

Although LMMSE estimation has low complexity, the second order channel
statistics of every hop are often not available at all nodes.
Therefore, ML is studied in this paper to provide potential
solutions to this problem by assuming the radio channels to be
\emph{deterministic}. To apply ML, the probability density function (pdf)
of $\mathbf{z}_{3}$ can be written as
\begin{flalign}\label{(29_0)}
&p\left(\mathbf{z}_{3}|\mathbf{\theta}\right)=\pi^{-N}|\mathbf{R}_{\tilde{\mathbf{n}}}|^{-1}
\textup{exp}\big(-\left(\mathbf{z}_{3}-\alpha_{1}\tilde{\alpha}_{3}\mathbf{T\Lambda
\theta}\right)^{H}&\nonumber\\
&\mathbf{R}_{\tilde{\mathbf{n}}}^{-1}
\left(\mathbf{z}_{3}-\alpha_{1}\tilde{\alpha}_{3}\mathbf{T\Lambda
\theta}\right)\big).&
\end{flalign}

The log-likelihood function is thus given by
\begin{flalign}\label{(29)}
&\textup{log}p\left(\mathbf{z}_{3}|\boldsymbol{\theta}\right)=-\left(\mathbf{z}_{3}-\alpha_{1}\tilde{\alpha}_{3}\mathbf{T\Lambda
\boldsymbol{\theta}}\right)^{H}&\nonumber\\
&\mathbf{R}_{\tilde{\mathbf{n}}}^{-1}\left(\mathbf{z}_{3}-\alpha_{1}\tilde{\alpha}_{3}\mathbf{T\Lambda
\boldsymbol{\theta}}\right)
-\textup{log}\left(|\mathbf{R}_{\tilde{\mathbf{n}}}|\right)-N\textup{log}\left(\pi\right).&
\end{flalign}

By maximizing \eqref{(29)}, ML estimates of
$\theta_{1}$ and $\theta_{2}$ can be obtained from
\begin{flalign}\label{estimations}
&\{\hat{\theta_{1}}, \hat{\theta_{2}}\} =\arg\;\underset{\theta_{1},
\theta_{2}}{\textup{min}}
\left(\mathbf{z}_{3}-\alpha_{1}^2\tilde{\alpha}_{3}\theta_{1}\mathbf{t}_{1}-\alpha_{1}\alpha_{2}
\tilde{\alpha}_{3}\theta_{2}\mathbf{t}_{2}\right)^{H}&\nonumber\\
&\mathbf{R}_{\tilde{\mathbf{n}}}^{-1}
\big(\mathbf{z}_{3}-\alpha_{1}^2\tilde{\alpha}_{3}\theta_{1}\mathbf{t}_{1}-\alpha_{1}\alpha_{2}\tilde{\alpha}_{3}\theta_{2}\mathbf{t}_{2})
+\log\left(|\mathbf{R}_{\tilde{\mathbf{n}}}|\right).&
\end{flalign}

$\mathbf{R}_{\tilde{\mathbf{n}}}$ can be expressed explicitly as
\begin{flalign}
&\mathbf{R}_{\tilde{\mathbf{n}}}=\sigma_{n}^{2}\big(\xi\mathbf{I}_{N}+\alpha_{1}^{2}
\tilde{\alpha_{3}}|\theta_{1}|\varepsilon\mathbf{T}\left(\mathbf{T}^{H}\mathbf{T}\right)^{-1}
\mathbf{T}^{H}\big)&\nonumber\\
&=\sigma_{n}^{2}\xi\left(\mathbf{I}_{N}+a\mathbf{P}_{T}\right),&
\end{flalign}
where $\mathbf{P}_{T}$ is the projection matrix of the space spanned
by the training matrix $\mathbf{T}$, and
$a=\frac{\alpha_{1}^{2}\tilde{\alpha}_{3}^{2}\varepsilon}{\xi}|\theta_{1}|$.
As the estimate $\hat{\theta}_{2}$ is independent of the second
term in (34), we can obtain $\hat{\theta}_{2}$ simply from the
least-squares approach under a given $\theta_{1}$ as
\begin{flalign}\label{29-2}
&\hat{\theta}_{2}=\textup{arg}\;\underset{\theta_{2}}{\textup{min}}\;
\big(-\left(\mathbf{z}_{3}-\alpha_{1}^2\tilde{\alpha}_{3}\theta_{1}\mathbf{t}_{1}-\alpha_{1}\alpha_{2}
\tilde{\alpha}_{3}\theta_{2}\mathbf{t}_{2}\right)^{H}&\nonumber\\
&\mathbf{R}_{\tilde{\mathbf{n}}}^{-1}
\big(\mathbf{z}_{3}-\alpha_{1}^2\tilde{\alpha}_{3}\theta_{1}\mathbf{t}_{1}-\alpha_{1}\alpha_{2}
\tilde{\alpha}_{3}\theta_{2}\mathbf{t}_{2})\big).&
\end{flalign}

Consequently, $\hat{\theta_{2}}$ can be written as

\begin{equation}
\hat{\theta}_{2}=\frac{\mathbf{t}_{2}^{H}\mathbf{R}_{\tilde{\mathbf{n}}}^{-1}}{\alpha_{1}
\alpha_{2}\tilde{\alpha}_{3}\mathbf{t}_{2}^{H}\mathbf{R}_{\tilde{\mathbf{n}}}^{-1}
\mathbf{t}_{2}}\left(\mathbf{z}_{3}-\alpha_{1}^{2}\alpha_{3}\theta_{1}\mathbf{t}_{1}\right).
\end{equation}

As the inverse of the equivalent noise is
$\mathbf{R}_{\tilde{\mathbf{n}}}^{-1}=\frac{1}{\sigma_{n}^{2}\xi}\left(\mathbf{I}-\frac{a}{1+a}\mathbf{P}_{T}\right)$,
$\hat{\theta_{2}}$ can be further simplified to
\begin{equation}
\hat{\theta}_{2}=\frac{\mathbf{t}_{2}^{H}}{\alpha_{1}\alpha_{2}\tilde{\alpha}_{3}Q_{2}}
\left(\mathbf{z}_{3}-\alpha_{1}^{2}\alpha_{3}\theta_{1}\mathbf{t}_{1}\right).
\end{equation}

By substituting $\hat{\theta_{2}}$ back into (34), the log-likelihood
function can be reformulated as
\begin{align}\label{57}
&\hat{\theta}_{1}=\arg \;\underset{h}{\textup{min}}\;
\bigg\{\mathbf{z}_{3}^{H}\mathbf{B}\mathbf{z}_{3}-2\alpha_{1}^{2}\tilde{\alpha}_{3}\mathfrak{R}\left\{\theta_{1}\mathbf{z}_{3}^{H}
\mathbf{B}\mathbf{z}_{3}\right\}&\nonumber\\
&+\alpha_{1}^{4}\alpha_{3}^{2}
|\theta_{1}|^{2}\mathbf{t}_{1}^{H}\mathbf{B}\mathbf{t}_{1}+\textup{log}\left(|\mathbf{R}_{\tilde{\mathbf{n}}}|\right)\bigg\},&
\end{align}
where
$\mathbf{A}=\mathbf{I}-\frac{\mathbf{t}_{2}\mathbf{t}_{2}^{H}\mathbf{R}_{\tilde{\mathbf{n}}}^{-1}}
{\mathbf{t}_{2}^{H}\mathbf{R}_{\tilde{\mathbf{n}}}^{-1}\mathbf{t}_{2}}$,
and
\begin{equation}\label{58}
\mathbf{B}\triangleq\mathbf{A}^{H}\mathbf{R}_{\tilde{\mathbf{n}}}^{-1}\mathbf{A}
=\left(\mathbf{R}_{\tilde{\mathbf{n}}}^{-1}-\frac{\mathbf{R}_{\tilde{\mathbf{n}}}^{-1}\mathbf{t}_{2}
\mathbf{t}_{2}^{H}\mathbf{R}_{\tilde{\mathbf{n}}}^{-1}}
{\mathbf{t}_{2}^{H}\mathbf{R}_{\tilde{\mathbf{n}}}^{-1}\mathbf{t}_{2}}\right).
\end{equation}

Since $\mathbf{B}$ and the determinant of
$\mathbf{R}_{\tilde{\mathbf{n}}}$ contain only the amplitude
$|\theta_{1}|$, the phase $\angle{\theta_{1}}$ of $\theta_1$ can be
independently estimated as
\begin{equation}\label{59}
\angle{\hat{\theta}_{1}}=-\angle{\left(\mathbf{z}_{3}^{H}\mathbf{B}\mathbf{t}_{1}\right)}.
\end{equation}

By minimizing the expression in \eqref{estimations}, $|\theta_{1}|$
can be estimated as
\begin{flalign}\label{60}
&\widehat{|\theta_{1}|}=&\nonumber\\
&\textup{arg}\;\underset{|\theta_{1}|}{\textup{min}}\;
\underbrace{\mathbf{z}_{3}^{H}\mathbf{B}\mathbf{z}_{3}-2\alpha_{1}^{2}\tilde{\alpha}_{3}|\theta_{1}|
|\mathbf{z}_{3}^{H}\mathbf{B}\mathbf{t}_{1}|+\alpha_{1}^{4}\tilde{\alpha}_{3}^{2}|\theta_{1}|^{2}
\mathbf{t}_{1}^{H}\mathbf{B}\mathbf{t}_{1}}_{f_{1}(a)}+&\nonumber\\
&\textup{log}\left(|\mathbf{R}_{\tilde{\mathbf{n}}}|\right),&\\
&=\textup{arg}\;\underset{|\theta_{1}|}{\textup{min}}\
\underbrace{\mathbf{z}_{3}^{H}\mathbf{B}\mathbf{z}_{3}-2\frac{\alpha_{1}^{2}
\tilde{\alpha}_{3}a}{a_{0}}|\mathbf{z}_{3}^{H}\mathbf{B}\mathbf{t}_{1}|
+\left(\frac{\alpha_{1}^{2}\tilde{\alpha}_{3}a}{a_{0}}\right)^{2}\mathbf{t}_{1}^{H}
\mathbf{B}\mathbf{t}_{1}}_{f_{1}(a)}+&\nonumber\\
&\textup{log}\left(|\mathbf{R}_{\tilde{\mathbf{n}}}|\right).&\label{61}
\end{flalign}

Considering that $a=a_{0}|\theta_{1}|$ is a simple multiple factor
of the estimate $|\theta_{1}|$ with
$a_{0}=\frac{\alpha_{1}^{2}\tilde{\alpha}_{3}^{2}\varepsilon}{\xi}$,
the minimization of \eqref{60} in terms of $|\theta_{1}|$ can be
solved via searching the corresponding $a$. Then, the derivative of
the first part $f_{1}(a)$ in \eqref{61} with respect to $a$ can be
expressed as
\begin{flalign}\label{f_1a}
&\dot{f}_{1}(a)=\frac{\partial f_{1}(a)}{\partial
a}=\mathbf{z}_{3}^{H}\frac{\partial \mathbf{B}}{\partial
a}\mathbf{z}_{3}-2\frac{\alpha_{1}^{2}\tilde{\alpha}_{3}}{a_{0}}
\frac{\partial
\left(a|\mathbf{z}_{3}^{H}\mathbf{B}\mathbf{t}_{1}|\right)}
{\partial a}&\nonumber\\
&+\left(\frac{\alpha_{1}^{2}\tilde{\alpha}_{3}}{a_{0}}\right)^{2}\mathbf{t}_{1}^{H}\frac{\partial
\left(a^{2}\mathbf{B}\right)}{\partial a}\mathbf{t}_{1}.&
\end{flalign}

According to the definition of the matrix $\mathbf{B}$ in \eqref{58},
the derivative of $\mathbf{B}$ with respect to $a$ can be
expressed as
\begin{equation}\label{(62)}
\frac{\partial \mathbf{B}}{\partial
a}=\frac{1}{\sigma_{n}^{2}\xi\left(1+a\right)^{2}}\left(-\mathbf{P}_{T}+\frac{1}{Q_{2}}\mathbf{t}_{2}\mathbf{t}_{2}^{H}\right).
\end{equation}

In order to obtain an analytical solution for $a$, the specific form
of $|\mathbf{z}_{3}^{H}\mathbf{B}\mathbf{t}_{1}|$ in \eqref{f_1a}
can be rewritten as
\begin{equation}\label{zbt}
|\mathbf{z}_{3}^{H}\mathbf{B}\mathbf{t}_{1}|=\frac{1}{\sigma_{n}^{2}\xi\left(1+a\right)}
|\mathbf{z}_{3}^{H}\mathbf{t}_{1}-\rho^{*}\sqrt{Q_{1}/Q_{2}}\mathbf{z}_{3}^{H}\mathbf{t}_{2}|.
\end{equation}

The estimated phase $\widehat{\angle{\theta_{1}}}$ for
$|\mathbf{z}_{3}^{H}\mathbf{B}\mathbf{t}_{1}|$ in \eqref{zbt} is
re-expressed as
\begin{equation}
\widehat{\angle{\theta_{1}}}=-\angle{\left(\mathbf{z}_{3}^{H}\mathbf{B}\mathbf{t}_{1}\right)}
=-\angle{\left(\mathbf{z}_{3}^{H}\mathbf{t}_{1}-\rho^{*}\sqrt{Q_{1}/Q_{2}}\mathbf{z}_{3}^{H}\mathbf{t}_{2}\right)},
\end{equation}
which indicates that $\widehat{\angle{\theta_{1}}}$ is
independent of $|\theta_{1}|$.

Similarly, the derivation of the other items of $\dot{f}_{1}(a)$ in
\eqref{f_1a} is directly given by
$\mathbf{t}_{1}^{H}\mathbf{B}\mathbf{t}_{1}=\frac{1-\rho^{2}}{1+a}Q_{1}$,
and $\mathbf{z}_{3}^{H}\frac{\partial \mathbf{B}}{\partial
a}\mathbf{z}_{3}=\frac{1}{\sigma_{n}^{2}\xi\left(1+a\right)^{2}}\left(-\mathbf{z}_{3}^{H}\mathbf{P}_{T}\mathbf{z}_{3}+\frac{1}{Q_{2}}
|\mathbf{z}_{3}^{H}\mathbf{t}_{2}|^{2}\right)$. Therefore, the
derivative of $f_{1}(a)$ can be re-organized as
\begin{equation}
\begin{split}
\dot{f}_{1}(a)&=\frac{1}{\sigma_{n}^{2}\xi(1+a)^{2}}\bigg[\left(-\mathbf{z}_{3}^{H}
\mathbf{P}_{T}\mathbf{z}_{3}+\frac{1}{Q_{2}}|\mathbf{z}_{3}^{H}\mathbf{t}_{2}|^{2}\right)\\
&-2\frac{\alpha_{1}^{2}\tilde{\alpha}_{3}}{a_{0}}|\mathbf{z}_{3}^{H}\mathbf{t}_{1}-\rho^{*}
\sqrt{Q_{1}/Q_{2}}\mathbf{z}_{3}^{H}\mathbf{t}_{2}|\\
&+\frac{\alpha_{1}^{4}\tilde{\alpha}_{3}^{2}}{a_{0}^{2}}(2a+a^{2})(1-\rho^{2})Q_{1}\bigg].
\end{split}
\end{equation}

The remaining part of (43) is to calculate the derivative of
$\textup{log}\left(|\mathbf{R}_{\tilde{\mathbf{n}}}|\right)$. Since
$\mathbf{P}_{T}$ is an orthogonal projection matrix, both
$\mathbf{I}$ and $\mathbf{P}_{T}$ are symmetric matrices. We can
define
$f_{2}(a)\!=\!\textup{log}\,\textup{det}\left(\mathbf{I}+a\mathbf{P}_{T}\right)$,
which can be transformed into
\begin{equation}
f_{2}(a)=\sideset{}{_{i=1}^{N}}\sum\textup{log}\left(1+a\lambda_{i}\right),
\end{equation}
where $\lambda_{i}$ is the eigenvalue of the projection matrix
$\mathbf{P}_{T}$.

The derivative of
$\textup{log}\left(|\mathbf{R}_{\tilde{\mathbf{n}}}\right|)$ with
respect to $a$ is equal to $\dot{f}_{2}(a)$, which can be expressed
as
\begin{equation}
\dot{f}_{2}(a)=\sideset{}{_{i=1}^{N}}\sum\frac{\lambda_{i}}{1+a\lambda_{i}}.
\end{equation}

The eigenvalue projection matrix $\mathbf{P}_{T}$ is only $0$ or
$1$, and the number of such that $\lambda_{i}=1$ is equal to the rank of
$\mathbf{P}_{T}$. The projection matrix satisfies
$\textup{tr}\{\mathbf{P}_{T}\}=\textup{rank}\left(\mathbf{P}_{T}\right)$.
Therefore, $\dot{f}_{2}(a)$ is given by
\begin{equation}\label{(67)}
\dot{f}_{2}(a)=\frac{r}{1+a},
\end{equation}
where $r=\textup{tr}\{\mathbf{P}_{T}\}$. Denoting by $f(a)$
the objective function in (43) and substituting the
derivatives obtained in (51) into $f(a)$, the derivative of $f(a)$
can be expressed concisely as
\begin{equation}
\dot{f}(a)=\frac{C_{1}a^{2}+C_{2}a+C_{3}}{\sigma_{n}^{2}\xi
a_{0}^{2}\left(1+a\right)^{2}},
\end{equation}
where
$C_{1}=\alpha_{1}^{4}\tilde{\alpha}_{3}^{2}\left(1-|\rho|^{2}\right)Q_{1}$,
$C_{2}=2\alpha_{1}^{4}\tilde{\alpha}_{3}^{2}\left(1-|\rho|^{2}\right)Q_{1}$\\
$+ ra_{0}^{2}\sigma_{n}^{2}\xi$, and
$C_{3}=a_{0}^{2}\left(-\mathbf{z}_{3}^{H}\mathbf{P}_{T}\mathbf{z}_{3}+\frac{1}{Q_{2}}|
\mathbf{z}_{3}^{H}\mathbf{t}_{2}|^{2}\right)-2\alpha_{1}^{2}
\tilde{\alpha}_{3}a_{0}|\mathbf{z}_{3}^{H}\mathbf{t}_{1}
-\rho^{*}\sqrt{Q_{1}/Q_{2}}\mathbf{z}_{3}^{H}\mathbf{t}_{2}|+ra_{0}^{2}\sigma_{n}^{2}\xi$.

Depending on the value of $|\rho|$, the coefficient $C_{1}$ is
larger than or equal to zero. The solution to the above equation can
be obtained in the following two cases.

\textbf{Case I:} When $|\rho|=1$, $\mathbf{t}_{1}$ and
$\mathbf{t}_{2}$ are fully correlated and $C_{1}=0$, the solution
can be obtained as $a=-\frac{C_{3}}{C_{2}}$ is straightforward. If
the coefficient $C_{2}$ is larger than zero owing to $r>0$, the
solution $a=-\frac{C_{3}}{C_{2}}$ is the global minimum of $f(a)$.
Considering that $a\geq0$, the estimate of $a$ is given by
\begin{equation}
\hat{a}=\textup{max}\{-C_{3}/C_{2}, 0\}.
\end{equation}

In practice, designing fully correlated training is inadvisable
since the two channel parameters $\theta_{1}$ and $\theta_{2}$ would
then be indistinguishable.

\textbf{Case II:} When $|\rho|<1$ and $C_{1}>0$, $\mathbf{t}_{1}$
and $\mathbf{t}_{2}$ are partially correlated or even orthogonal.
The root of the quadratic function $\dot{f}(a)$ is determined by the
discriminant $C_{2}^{2}-4C_{1}C_{3}$. When
$C_{2}^{2}-4C_{1}C_{3}\geq0$, the two roots of $\dot{f}(a)$ can be
written in
$a_{1,2}=\frac{-C_{2}\pm\sqrt{C_{2}^{2}-4C_{1}C_{3}}}{2C_{1}}$. It
is readily affirmed that
$a_{1}=\frac{-C_{2}-\sqrt{C_{2}^{2}-4C_{1}C_{3}}}{2C_{1}}$ must be
a local maximum, and
$a_{2}=\frac{-C_{2}+\sqrt{C_{2}^{2}-4C_{1}C_{3}}}{2C_{1}}$
 must be a local minimum due to the fact that $C_{1}>0$.
Considering $a\geq0$, the estimate of $a$ is simply expressed as
\begin{equation}
\hat{a}=\textup{max}\left\{\frac{-C_{2}+\sqrt{C_{2}^{2}-4C_{1}C_{3}}}{2C_{1}},
0\right\}.
\end{equation}

When the discriminant $C_{2}^{2}-4C_{1}C_{3}<0$, $\dot{f}(a)$ has no
roots and $f(a)$ is a monotonically increasing function. As a
consequence, the estimate is simply $\hat{a}=0$ where
the minimum of $f(a)$ lies. Once the estimate of $a$ is
obtained, the corresponding $|\theta_{1}|$ can be calculated from
$|\theta_{1}|=a/a_{0}$.

\section{Training Design for $4-$Hop WCNs}

In this section, based on the channel estimation schemes proposed in
Section III, the corresponding optimal training designs obtained by
minimizing MSE for LMMSE estimation are characterized in the following lemmas
and proposition.

\begin{lemma}\label{lemma1}
\emph{The optimal LMMSE-based training in $4-$hop WCNs is orthogonal
with maximum allowable transmit power.}
\end{lemma}

\begin{IEEEproof}
See Appendix A.
\end{IEEEproof}

Due to the nonlinearity of ML, it is difficult to obtain a closed-form
expression of the corresponding MSE, and the training design method for
LMMSE estimation based on minimizing MSE is not suitable. Instead, we resort to
designing training sequences from minimizing the Cram\'{e}r Rao Lower
Bound (CRLB), which provides a lower bound on the variance of unbiased
estimates.

\begin{proposition}\label{proposition1}
The CRLBs for $\boldsymbol{\theta}$ in $4-$hop WCNs are
\begin{align}
&\textup{CRLB}_{\theta_{1}}=\frac{D_{3}(D_{1}D_{3}-|D_{2}|^{2})}{|D_{2}|^{4}-2
D_{1}D_{3}|D_{2}|^{2}+D_{1}^{2}D_{3}^{2}-|D_{4}|^{2}D_{3}^{2}},\label{CRLB_1}\\
&\textup{CRLB}_{\theta_{2}}=\frac{(-D_{1}|D_{2}|^{2}-|D_{4}|^{2}D_{3}+D_{1}^{2}D_{3})}
{|D_{2}|^{4}-2D_{1}D_{3}|D_{2}|^{2}+D_{1}^{2}D_{3}^{2}-|D_{4}|^{2}D_{3}^{2}},\label{CRLB_2}
\end{align}
where
\begin{align}
&D_{1}=\frac{\alpha_{1}^{4}\tilde{\alpha}_{3}^{2}Q_{1}}{\sigma_{n}^{2}\xi(1+a)}
+\frac{a_{0}^{2}(r-2)^{2}}{4(1+a)^{2}},\quad
D_{2}=\frac{\alpha_{1}^{4}\tilde{\alpha}_{3}^{2}\rho\sqrt{Q_{1}Q_{2}}}{\sigma_{n}^{2}\xi(1+a)},\nonumber\\
&D_{3}=\frac{\alpha_{1}^{4}\tilde{\alpha}_{3}^{2}Q_{2}}{\sigma_{n}^{2}\xi(1+a)},
\quad
D_{4}=\frac{a_{0}^{2}(r-2)^{2}\theta_{1}^{2}}{4(1+a)^{2}|\theta_{1}|^{2}}.\nonumber
\end{align}
\end{proposition}

\begin{IEEEproof}
See Appendix B.
\end{IEEEproof}

Once the explicit forms of $\textup{CRLB}_{\theta_{1}}$ and
$\textup{CRLB}_{\theta_{2}}$ are obtained, optimal training
design by minimizing CRLBs can be developed according to the
following lemma.
\begin{lemma}
\emph{The optimal training design based on minimizing the CRLB is orthogonal
with maximum allowable transmit power.}
\end{lemma}

\begin{IEEEproof}
See Appendix C.
\end{IEEEproof}

\section{General Network Coded $2N-$Hop WCNs}

In this section, the network coding transmission strategy, radio
channel estimation and corresponding training design schemes for
general MH-WCNs are considered as extensions of $4-$hop
WCNs. The $2N-$hop WCNs are bidirectional multi-hop networks
with $2N-1$ cascaded cooperative intermediate nodes
$\mathbb{R}_{2n-1}, n=1,\ldots,N (N \geq 2)$ and two desired nodes
$\mathbb{T}_{1}$ and $\mathbb{T}_{2}$. The average transmission power of
$\mathbb{T}_{i} (i = 1, 2)$ and $\mathbb{R}_{i} (i = 1, 2, ..., 2N -
1)$ are set as $P_{i}$ and $P_{ri}$, respectively. The radio
channels are reciprocal as time-division-duplex (TDD) is utilized.
Denote the radio channel gain from $\mathbb{T}_{1}$ to $\mathbb{R}_{1}$
as $h_{1}$, from $\mathbb{T}_{2}$ to $\mathbb{R}_{2}$ as $g_{1}$,
from $\mathbb{R}_{2N-3}$ to $\mathbb{R}_{2N-1}$ as $h_{N}$, and from
$\mathbb{R}_{2N-4}$ to $\mathbb{R}_{2N-2}$ as $g_{N}$. In addition,
the radio channel gain between $\mathbb{R}_{2i-3}$ and
$\mathbb{R}_{2i-1}$ is denoted by $h_{i}$, and that between
$\mathbb{R}_{2i-2}$ and $\mathbb{R}_{2i}$ is denoted by $g_{i}$. All
radio channels satisfy $h_{i}\sim\mathcal{C}\mathcal{N}(0,
\sigma_{2i-1}^{2})$, and $g_{i}\sim\mathcal{C}\mathcal{N}(0,
\sigma_{2i}^{2}), i=1, \ldots, N$.

\subsection{Network Coding Transmission Strategy in $2N-$Hop WCNs}

Analogously to the network coding transmission strategy in $4-$hop WCNs,
the signals $x_{1}^{(1)}$ and $x_{2}^{(1)}$ transmitted by two
desired nodes in $2N-$hop WCNs are exchanged in $2N$ time phases,
with nearly every node required to eliminate self-interference.
The signal received at $\mathbb{T}_{1}$ is
\begin{equation}\label{T2_X2}
\begin{split}
d_{1}&=\alpha_{1}h_{1}s_{1}^{(N)}+\alpha_{1}\alpha_{3}h_{1}h_{2}s_{3}^{(N-1)}+\ldots+\\
&\left(\sideset{}{_{i=1}^{N}}\prod\alpha_{2i-1}h_{i}\right)s_{2N-1}^{(1)}+n_{1}\\
&=\left[\sideset{}{_{j=1}^{N}}\sum\left(\sideset{}{_{i=1}^{j}}\prod\alpha_{2i-1}
h_{i}\right)s_{2j-1}^{(N+1-j)}\right]+n_{1}\end{split}.
\end{equation}

Similarly, the signal received at $\mathbb{T}_{2}$ is given by
\begin{equation}\label{(58)}
\begin{split}
d_{2} = \left[\sideset{}{_{j=1}^{N-1}}\sum\left(\sideset{}{_{i=1}^{j}}\prod\alpha_{2i}g_{i}\right)s_{2j}^{(N+1-j)}\right]\\
+\alpha_{2N-1}\left(\sideset{}{^{N-1}_{i=1}}\prod\alpha_{2i}g_{i}\right)g_{N}s_{2N-1}^{(1)}+n_{2}\end{split},
\end{equation}
where
\begin{equation*}
\begin{split}
s_{2j-1}^{(N+1-j)}&\!=\left(\sideset{}{^{j-1}_{p=1}}\prod\alpha_{2p-1}h_{p+1}\right)
h_{1}x_{1}^{(N+1-j)} + n_{2j-1}^{(N+1-j)}\\
&+\left[\sideset{}{_{p=1}^{j-1}}\sum\left(\sideset{}{_{q=p}^{j-1}}\prod\alpha_{2q-1}
h_{q+1}\right)n_{2p-1}^{(N+1-j)}\right],\\
s_{2j}^{N+1-j}&\!=\left(\sideset{}{_{p=1}^{j-1}}\prod\alpha_{2p}g_{p+1}\right)g_{1}
x_{2}^{(N+1-j)} + n_{2j}^{(N+1-j)}\\
&+\left[\sideset{}{_{p=1}^{j-1}}\sum\left(\sideset{}{_{q=p}^{j-1}}
\prod\alpha_{2q}g_{q+1}\right)n_{2p}^{(N+1-j)}\right],
\end{split}
\end{equation*}
\begin{equation*}
\begin{split}
\alpha_{2i-1}&=\sqrt{\dfrac{P_{r\left(2i-1\right)}}{P_{r\left(2i-3\right)}\sigma_{2i-3}^{2}
+P_{r\left(2i+1\right)}\sigma_{2i+2}^{2}+\sigma_{n}^{2}}},\\
\alpha_{2i}&=\sqrt{\dfrac{P_{r\left(2i\right)}}{P_{r\left(2i-2\right)}\sigma_{2i-2}^{2}
+P_{r\left(2i+2\right)}\sigma_{2i+2}^{2}+\sigma_{n}^{2}}},\\
&\;i = 2, 3, \ldots, N-1, j = 2, \ldots, N-1,
\end{split}
\end{equation*}
\begin{equation*}
\begin{split}
&s_{1}^{(N)}=h_{1}x_{1}^{(N)}+n_{1}^{(N)}, \;s_{2}^{(N)}=g_{1}x_{2}^{(N)}+n_{2}^{(N)},\\
&s_{2N-1}^{(1)}=\alpha_{2N-3}h_{N}s_{2N-3}^{(1)}+\alpha_{2N-2}g_{N}s_{2N-2}^{(1)}+n_{2N-1}^{(1)}.
\end{split}
\end{equation*}

The target information for $\mathbb{T}_{1}$ in \eqref{T2_X2} is
$x_{2}^{(1)}$ transmitted from $\mathbb{T}_{2}$. The other terms are the
self-interference $x_{1}^{(1)}, x_{1}^{(2)},\ldots,x_{1}^{(N)}$
induced by the network coding transmission strategy and the Gaussian
noise generated by the nodes during the transmission. By designing
the training process meticulously, $x_{2}^{(1)}$ can be extracted at
$\mathbb{T}_{1}$ by the corresponding channel estimation scheme.
Moreover, not only does signal detection at the desired nodes
need radio channel estimation, but the other intermediate nodes
also need the corresponding radio channel coefficients to suppress
the self-interference and make the network coding transmission
strategy feasible. For instance, the received signals at the nodes
$\mathbb{R}_{2i\!-\!1},i\!=\!2,4,\ldots$ on the left-hand side of
$\mathbb{R}_{2N-1}$ when $\mathbb{T}_{1}$ receives $x_{2}^{(1)}$
are

\begin{flalign}
&d_{2i-1} = s_{2i-1}^{(N-i+2)}+\sideset{}{_{j=i+1}^{N}}\sum\left(\sideset{}{_{m=i+1}^{j}}\prod\alpha_{2m-1}h_{m}\right)&\nonumber\\
&s_{2j-1}^{(N-j+2)}+\alpha_{2j-3}\alpha_{2j-1}h_{j}^{2}&\\
&\left[\sideset{}{_{j=i+1}^{N}}\sum\left(\sideset{}{_{m=i+1}^{j}}\prod
\alpha_{2m-1}h_{m}\right)s_{2j-1}^{(N-j+1)}+s_{2i-1}^{(N-i+1)}\right],&\nonumber
\end{flalign}
where
$\sideset{}{_{j\!=\!i\!+\!1}^{N}}\sum\left(\sideset{}{_{m\!=\!i\!+\!1}^{j}}\prod
\alpha_{2m-1}h_{m}\right)s_{2j-1}^{(N-j+1)}\!+\!s_{2i-1}^{(N-i+1)}$
is the network coded signal that $\mathbb{R}_{2i\!-\!1}$ receives at
the previous round, which is the self-interference to be deleted by
$\mathbb{R}_{2i\!-\!1}$ through estimating $h_{j}^{2}$. Similarly,
the nodes $\mathbb{R}_{2i-2},i\!=\!2,4,\ldots$ located on the
right-hand side of $\mathbb{R}_{2N-1}$ should estimate $g_{j}^{2}$ for the
interference cancelation. Therefore, the training process design in
$2N$-hop WCNs should satisfy the above-mentioned requirements, which
is similar to the training design in 4-hop WCNs.

\subsection{Training Design in $2N-$Hop WCNs}

Like the $4-$hop scenario, the desired nodes $\mathbb{T}_{1}$
and $\mathbb{T}_{2}$ in $2N-$hop WCNs send $\mathbf{t}_{1}$ and
$\mathbf{t}_{2}$, and the intermediate node $\mathbb{R}_{2N-1}$
transmits $\mathbf{t}_{r}$, which is orthogonal to $\mathbf{t}_{1}$
and $\mathbf{t}_{2}$. The difference from 4-hop WCNs is that every
intermediate node needs to transmit $\mathbf{t}_{r}$ to the previous
node to acquire the corresponding radio channel coefficients for
self-interference suppression. Once $\mathbf{t}_{r}$ is back from
the previous node, $\mathbf{t}_{r}$ is deleted, and thus only the
remaining $\mathbf{t}_{1}$ or $\mathbf{t}_{2}$ is forward or
backward. For simplicity, the estimation error and residual noise
are assumed to be Gaussian. The received training
signal at $\mathbb{T}_{1}$ is
\begin{equation}
\begin{split}
\mathbf{z}_{1}^{(N)}&=\alpha_{1}h_{1}\mathbf{r}_{1}^{(N)}+\mathbf{n}_{\mathbf{z}}^{(N)}
=\sideset{}{_{i=1}^{N}}\prod\alpha_{2i-1}\mathbf{T}\mathbf{\Lambda}_{\mathbf{z}}\mathbf{h}_{z}+\tilde{\mathbf{n}}_{z},
\end{split}
\end{equation}
where
$\mathbf{\Lambda}_{\mathbf{z}}=\textup{diag}\{\sideset{}{_{i=1}^{N-1}}\prod\alpha_{2i-1},
\sideset{}{_{i=1}^{N-1}}\prod\alpha_{2i}\}$,
$\mathbf{h}_{z}=\textup{diag}\{\sideset{}{_{i=1}^{N}}\prod
h_{i}^{2}, \sideset{}{_{i=1}^{N}}\prod
h_{i}g_{i}\}=\textup{diag}\{\varpi_{1},\;\varpi_{2}\}$,
$\tilde{\mathbf{n}}_{2N-1}^{(1)}=\sideset{}{_{i=1}^{N-1}}\sum\big[\left(\sideset{}{_{j=i}^{N-1}}
\prod\alpha_{2j-1}h_{j+1}\right)\mathbf{n}_{2i-1}^{(1)}\big]
+\sideset{}{_{i=1}^{N-1}}\sum\big[\left(\sideset{}{_{j=i}^{N-1}}\prod\alpha_{2j}g_{j+1}\right)
\mathbf{n}_{2i}^{(1)}\big]+\mathbf{n}_{2N-1}^{(1)}$.

\begin{proposition}

When the number of communication nodes in $2N$-hop WCNs approaches
infinity, the MSE performance of $\varpi_{1}$ and $\varpi_{2}$
for the LMMSE scheme can be obtained respectively in certain
situations as
\begin{align}
&\sigma_{\varpi_{1}}^{2}=\frac{\sigma_{n}^{2}}{(1-\omega)(1-\rho^{2})Q_{1}},\nonumber\\
&\sigma_{\varpi_{2}}^{2}=\frac{\sigma_{n}^{2}}{(1-\omega)(1-\rho^{2})Q_{2}}.
\end{align}

Similarly, the CRLB performance of $\varpi_{1}$ and $\varpi_{2}$
for the ML scheme can be obtained respectively in certain situations
as
\begin{align}
&\textup{CRLB}_{\varpi_{1}}=\frac{\sigma_{n}^{2}}{(1-\omega)(1-\rho^{2})Q_{1}},\nonumber\\
&\textup{CRLB}_{\varpi_{2}}=\frac{\sigma_{n}^{2}}{(1-\omega)(1-\rho^{2})Q_{2}}.
\end{align}
\end{proposition}

\begin{IEEEproof}
See Appendix D.
\end{IEEEproof}

\section{Simulation Results}

In this section, we numerically evaluate the presented channel
estimation schemes along with the training design. The radio
channels $g_{i}$ and $h_{i}$ of every hop are assumed to be circularly
symmetric complex Gaussian random variables with zero means and unit
variances. The transmission power of the desired nodes and intermediate
nodes are assumed to be $P_{1} = P_{2} = P_{r1} = P_{r2} = P_{r3}$, and
the variances of the noise generated at the desired nodes and
intermediate nodes are assumed to be unity. The common
signal-to-noise ratio (SNR) is defined as $P_{1}/\sigma_{n}^{2} =
P_{1}$. The length of the training sequences $L$ is set as 8.  The
power of each symbol in the training sequences equals to $P_{1}$,
namely, the training power for $Q_{1}$ or $Q_{2}$ with the length of
$N$ is set as $NP_{1}$. The correlation coefficients $|\rho|=0$,
$|\rho|=0.5$, and $|\rho|=0.9$ as three examples for comparisons are
evaluated in simulations. Totally $10^{5}$ Monte-Carlo runs are
considered. Since the simulation results at $\mathbb{T}_{1}$ and
$\mathbb{T}_{2}$ are symmetric and exchangeable, the following
simulations are based on only the desired node
$\mathbb{T}_{1}$.

\subsection{LMMSE Channel Estimation Schemes}

The MSE performance for LMMSE estimation in $4-$hop WCNs under different
$|\rho|$'s is evaluated and compared in Fig. 4. LMMSE estimation achieves its best MSE
performance when $|\rho|\!=\!0$ for all values of
$\theta_{1}$ and $\theta_{2}$, which is consistent with
Lemma 1. The MSE gap increases when $|\rho|$ becomes large,
which indicates that LMMSE estimation is significantly
affected by the training structure and the optimal training design
is vital in $4-$hop MSNs. Moreover, the MSE discrepancy between
$\theta_{1}$ and $\theta_{2}$ indicates that LMMSE estimation is sensitive to the
statistical properties of the radio channel.

%\begin{figure}[!h]
%\centering
%\includegraphics[width=2.6in]{fig4.eps}
%\caption{LMMSE channel estimation MSEs versus SNR for $\theta_{1}$
%and $\theta_{2}$.} \label{fig4}\vspace*{-1em}
%\end{figure}

LMMSE estimation performance of $4-$hop WCNs in terms of the average effective
SNR (AESNR) is evaluated in Fig. 5, where the relationship between
AESNR and SNR is shown. Orthogonal training is used in
this simulation (i.e., $|\rho| = 0$). The dotted line is the
theoretical bound of the AESNR obtained by assuming that
perfect CSI is acquired at $\mathbb{T}_{1}$. The solid line is the
AESNR derived from the LMMSE method by taking the estimation errors
into account. When the SNR is low, the LMMSE AESNR is far below the
theoretical bound. As the SNR increases, the LMMSE AESNR
approaches the theoretical bound, which demonstrates that LMMSE estimation
is effective in $4-$hop MSNs.

\begin{figure}[!h]
\centering
\includegraphics[width=2.6in]{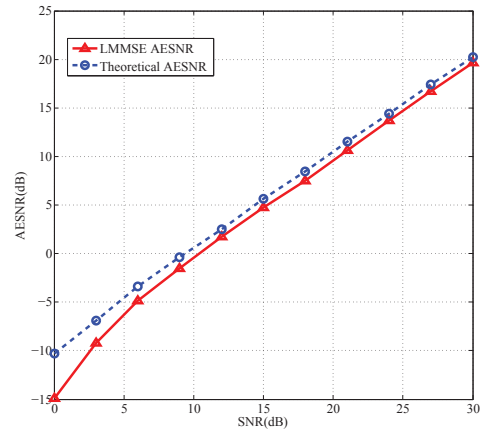}
\caption{Theoretical and LMMSE AESNR comparison with regards of
SNR.} \label{fig5}\vspace*{-1em}
\end{figure}

In Fig. 6, MSE performance is compared between the proposed LMMSE
and the baseline point-to-point estimation scheme. In the baseline
scheme, the training sequences are orthogonal to each other and the
feedback estimated CSI is perfectly known at the desired nodes. It
can be observed that the MSEs of $\theta_{1}$ and $\theta_{2}$ for
LMMSE estimation are lower than that for point-to-point estimation in the
region of high SNR, which demonstrates the effectiveness and
performance gains of LMMSE estimation.

%\begin{figure}[!h]
%\center
%\includegraphics[width=2.6in]{fig6.eps}
%\caption{MSE comparisons between LMMSE and traditional
%estimation.}\vspace*{-1em} \label{fig_6}
%\end{figure}

\subsection{ML Channel Estimation Schemes}

Similarly to LMMSE estimation, the MSE performance for $\theta_{1}$
and $\theta_{2}$ at different $|\rho|$'s when ML is utilized in
$4$-hop WCNs is shown in Fig. 7. It can be seen that ML achieves the
best performance when $|\rho| = 0$, which is the same as LMMSE
estimation. The MSE performance gaps increase with increasing SNR or
when $|\rho|$ is large, which resembles the results for LMMSE
estimation. The optimal training design can provide a significant
accuracy performance gain for ML in $4-$hop MSNs. In the low SNR
region, the MSE of $\theta_{2}$ is much larger than that of
$\theta_{1}$ because the estimation of $\theta_{2}$ is indirectly
obtained from the least-squares approach, which is based on the
estimated $\hat{\theta}_{1}$. On the other hand, the MSE performance
differences between $\theta_{1}$ and $\theta_{2}$ almost vanish in
the high SNR region due to the increasing estimation precision of
$\theta_{1}$.
%\begin{figure}[!h]
%\centering
%\includegraphics[width=2.6in]{fig7.eps}
%\caption{ML channel estimation MSEs versus SNR for $\theta_{1}$ and
%$\theta_{2}$.} \label{fig7}\vspace*{-1em}
%\end{figure}

The CRLBs of $\theta_{1}$ and $\theta_{2}$ versus SNR under
different $|\rho|$'s in $4-$hop WCNs are shown in Fig. 8. In the
case of $|\rho| = 0$, the CRLBs of $\theta_{1}$ and $\theta_{2}$
achieve the best performance, which validates Lemma 3.
Meanwhile, the optimal CRLB-based training structure coincides with
optimal orthogonal training derived from ML and LMMSE estimation in the
simulation as well as theoretical results.
%
%\begin{figure}[!h]
%\centering
%\includegraphics[width=2.6in]{fig8.eps}
%\caption{CRLB versus SNR for $\theta_{1}$ and $\theta_{2}$.}
%\label{fig8}\vspace*{-1em}
%\end{figure}

\subsection{Performance Evaluations for General $2N-$Hop WCNs}

When the $4$-hop case is extended to the general $2N-$hop case with
$2N-1$ intermediate nodes and two desired nodes, the MSE and CRLB
performance versus $N$ for LMMSE and ML estimation is compared in
Fig. 9, where SNR is set at $0$ dB, and $|\rho|$ is varied. The MSE
performance of LMMSE estimation is lower than the CRLB performance
of ML estimation when $N$ is small, which indicates that LMMSE
estimation often outperforms the unbiased ML estimation as it is
derived to solely minimize the MSE. As $N$ increases, the
performance of the unbiased estimator approaches the LMMSE estimator
and the MSE curves in Fig. 9 overlap with the CRLB curves under
different $|\rho|$'s after a small number of relays, which
corroborates with the analytical results in Section V. In addition,
the MSE and CRLB performance under different $|\rho|$'s is shown in
Fig. 9 as well. Similarly to results for $4-$hop WCNs, the
orthogonal training can achieve the best performance in $2N-$hop
MSNs, which indicates that the extension of the optimal orthogonal
training design from $4-$hop to $2N-$hop WCNs is reasonable and
effective.

%\begin{figure}[!h]
%\centering
%\includegraphics[width=2.6in]{fig9.eps}
%\caption{MSE/CRLB performance comparisons in terms of $N$.}
%\label{fig9}\vspace*{-1em}
%\end{figure}

In Fig. 10, MSE and CRLB performance is compared between
Monte-Carlo simulation and theoretical analysis, where $N$ is set at
$8$ and the value of $|\rho|$ is assumed to be $0$. In the low SNR
region, MSE and CRLB performance in the simulation are slightly
worse than that of the theoretical analysis because the noise is
approximated during the theoretical derivation. As SNR increases,
the impact of noise diminishes, and performance of the simulated
CRLB and MSE is close to that of the theoretical analysis.

\begin{figure}[!h]
\centering \includegraphics[width=2.6in]{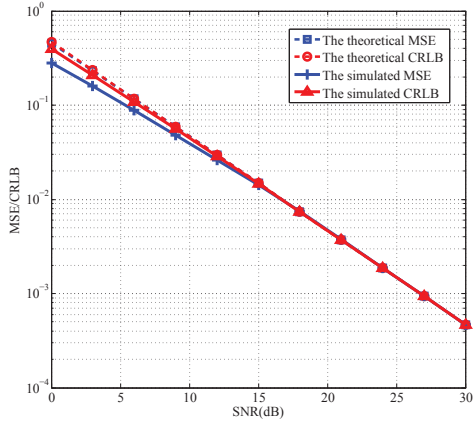}
\caption{MSE/CRLB comparisons between analysis and simulation.}
\label{fig10}\vspace*{-1em}
\end{figure}

\section{CONCLUSIONS}

Motivated by the socially enabled user cooperation and multi-hop
transmission in mobile sensor networks, the multi-hop
wireless communication networks as the key communication
technological component of MSNs have been studied in this paper.
A bidirectional network coding multi-hop transmission strategy has
been proposed, where the $4-$hop WCNs has been studied first as the
paradigm, followed by $2N-$hop WCNs. To implement network coded
MH-WCNs and improve transmission quality, radio channel
estimation and the corresponding training design have been
studied. Particularly, the LMMSE and ML radio channel estimation
methods have been proposed in $4-$hop WCNs firstly for the
acquisition of composite channel coefficients at desired nodes. A
closed-form MSE performance expression for the proposed LMMSE estimation
scheme has been derived, and an optimal training scheme has been designed to minimize
MSE. Due to the nonlinearity of ML radio channel estimation, design
of the training sequences via minimizing CRLB has been exploited.
Orthogonal training with the maximum allowable transmit power has
been proved to be optimal for both CRLB and MSE performance
criteria. According to the numerical and simulation results, both
LMMSE and ML estimators have demonstrated effectiveness to improve the
estimation accuracy in $4-$hop WCNs. Meanwhile, the extension of the
optimal orthogonal training design from $4-$hop WCNs to $2N-$hop
WCNs has been shown to be reasonable and effective. This work
improves the spectral efficiency and transmission quality of
MH-WCNs, which can promote the successful use of MSNs in
the future.

\appendices
\section{Proof of Lemma 1}

The resulting MSE for the LMMSE scheme can be derived from \cite{26}
as

\begin{equation}\label{Re_MSE}
\sigma_{\mathbf{\theta}}^{2}=\textup{tr}\big\{\left(\mathbf{R}_{\mathbf{\theta}}^{-1}
+\alpha_{1}^{2}\tilde{\alpha}_{3}^{2}\mathbf{\Lambda}\mathbf{T}^{H}\mathbf{R}_{\tilde{\mathbf{n}}}^{-1}
\mathbf{T}\mathbf{\Lambda}\right)^{-1}\big\}.
\end{equation}

The MSE matrix in \eqref{Re_MSE} can be expressed as
\begin{equation}\label{RE_MSE1}
\begin{split}
&\left(\mathbf{R}_{\mathbf{\theta}}^{-1}+\alpha_{1}^{2}\tilde{\alpha}_{3}^{2}\mathbf{\Lambda}\mathbf{T}^{H}
\mathbf{R}_{\tilde{\mathbf{n}}}^{-1}\mathbf{T}\mathbf{\Lambda}\right)^{-1}\\
&=\frac{1}{\lambda}\begin{pmatrix} \frac{1}{\sigma_{\theta_{2}}^{2}}+\nu\alpha_{2}^{2}\left[Q_{2}+a_{1}(1-\rho^{2})Q_{1}Q_{2}\right] \\
-\nu\alpha_{1}\alpha_{2}\left[\rho\sqrt{Q_{1}Q_{2}}+a_{3}(1-\rho^{2})Q_{1}Q_{2}\right] \\
-\nu\alpha_{1}\alpha_{2}\left[\rho^{*}\sqrt{Q_{1}Q_{2}}+a_{3}^{*}(1-\rho^{2})Q_{1}Q_{2}\right]
\frac{1}{\sigma_{\theta_{1}}^{2}}\\
+\nu\alpha_{1}^{2}\left[Q_{1}+a_{2}(1-\rho^{2})Q_{1}Q_{2}\right]\end{pmatrix},
\end{split}
\end{equation}
where
$a_{1}=\frac{\alpha_{1}^{2}\alpha_{3}^{2}}{\xi}\frac{\sigma_{1}^{2}\sigma_{3}^{2}\varepsilon}{\left(1-|\rho|^{2}\right)Q_{1}}$,
$a_{2}=\frac{\alpha_{1}^{2}\alpha_{3}^{2}}{\xi}\frac{\sigma_{1}^{2}\sigma_{3}^{2}\varepsilon}{\left(1-|\rho|^{2}\right)Q_{2}}$,
$a_{3}=A_{3}$,
$\nu=\frac{\alpha_{1}^{2}\alpha_{3}^{2}}{\tau^{*}\xi\sigma_{n}^{2}}$
are defined for notational simplicity. The parameter $\lambda$ can
be written as
\begin{equation}\label{RE_MSE_lamda}
\begin{split}
\lambda=\frac{1}{\sigma_{\theta_{1}}^{2}\sigma_{\theta_{2}}^{2}}+\nu^{2}(1-\rho^{2})Q_{1}Q_{2}\tau^{*}\\
+\frac{1}{\sigma_{\theta_{1}}^{2}}\kappa\alpha_{2}^{2}\left[Q_{2}+a_{1}(1-\rho^{2})Q_{1}Q_{2}\right]\\
+\frac{1}{\sigma_{\theta_{2}}^{2}}\nu\alpha_{1}^{2}\left[Q_{1}+a_{2}(1-\rho^{2})Q_{1}Q_{2}\right].
\end{split}
\end{equation}

To derive the expression in \eqref{RE_MSE_lamda}, the following
substitutions are defined: $x=1-\rho^{2}$,
$a=\frac{\alpha_{1}^{2}\tilde{\alpha}_{3}^{2}\sigma_{1}^{2}\sigma_{3}^{2}\varepsilon}{\xi}$,
$b=\frac{\alpha_{1}^{2}\tilde{\alpha}_{3}^{2}}{\sigma_{n}^{2}\xi}$.
The coefficients $a_{1}$, $a_{2}$, $a_{3}$ can be simplified as follows:
$a_{1}=\dfrac{a}{Q_{1}x}$,
$a_{2}=\dfrac{a}{Q_{2}x}$,
$a_{3}=-\dfrac{a\rho}{\sqrt{Q_{1}Q_{2}}x}$, respectively.
Substituting these simplified coefficients into $\tau^{*}$, after
re-organization, $\tau^{*}$ can be written as
\begin{equation}\label{(39)}
\tau^{*}=1+\frac{2a}{x}(1-\rho^{2})+a^{2}\frac{(1-\rho^{2})^{2}}{x^{2}}=(1+a)^{2}.
\end{equation}

Then, $\nu$ can be rewritten as $\nu=b/(1+a)^{2}$. Therefore, after
some algebra, the MSE performance
$\sigma_{\mathbf{\theta}}^{2}$ can be written as
\begin{equation}\label{(40)}
\sigma_{\mathbf{\theta}}^{2}=\frac{1}{\lambda\sigma_{\theta_{1}}^{2}\sigma_{\theta_{2}}^{2}}\left[\sigma_{\theta_{1}}^{2}+\sigma_{\theta_{2}}^{2}+\frac{b}{1+a}(\alpha_{2}^{2}Q_{2}+\alpha_{1}^{2}Q_{1})
\sigma_{\theta_{1}}^{2}\sigma_{\theta_{2}}^{2}\right].
\end{equation}

Further, $\lambda$ can be simplified as
\begin{equation}\label{(41)}
\begin{split}
&\lambda=\frac{1}{\sigma_{\theta_{1}}^{2}\sigma_{\theta_{2}}^{2}}+\frac{b}{(1+a)\sigma_{\theta_{1}}^{2}}\alpha_{2}^{2}Q_{2}\\
&+\frac{b}{(1+a)\sigma_{\theta_{2}}^{2}}\alpha_{1}^{2}
Q_{1}+\frac{b^{2}}{(1+a)^{2}}x\alpha_{1}^{2}\alpha_{2}^{2}Q_{1}Q_{2}.
\end{split}
\end{equation}

Since $\sigma_{\mathbf{\theta}}^{2}$ is a positive coefficient
independent of the correlation coefficient $\rho$, it is
monotonically decreasing with $\lambda$. Meanwhile, $\lambda$
monotonically increases with $x$ and decreases with $\rho$, and the MSE
$\sigma_{\mathbf{\theta}}^{2}$ can be proven to be proportional to
the correlation coefficient $\rho$ and achieves its minimum when
$\rho=0$, which means the best training should be orthogonal. Taking
the optimal value $|\rho|=0$, the MSE $\sigma_{\mathbf{\theta}}^{2}$ can
be rewritten as
\begin{equation}\label{(42)}
\sigma_{\mathbf{\theta}}^{2}=\frac{\sigma_{\theta_{1}}^{2}+\sigma_{\theta_{2}}^{2}+\frac{b}{1+a}
(\alpha_{2}^{2}Q_{2}+\alpha_{1}^{2}Q_{1})\sigma_{\theta_{1}}^{2}\sigma_{\theta_{2}}^{2}}{(1+\frac{b}{1+a}\alpha_{1}^{2}
\sigma_{\theta_{1}}^{2}Q_{1})(1+\frac{b}{1+a}\alpha_{2}^{2}\sigma_{\theta_{2}}^{2}Q_{2})}.
\end{equation}

The maximum training power assigned to $\mathbf{t}_{i}$ is assumed to
be $Q_{i}^{max}$, which is typically equal to $LP_{i}$ ($P_{i}$
is the transmission power at $\mathbb{T}_{i}$). Thus, the optimal power
can be allocated as $\|\mathbf{t}_{i}\|^{2}\in[0, Q_{i}^{max}]$ by
minimizing $\sigma_{\mathbf{\theta}}^{2}$. The derivatives of
$\sigma_{\mathbf{\theta}}^{2}$ with respect to $Q_{1}$ and $Q_{2}$
are given by
\begin{equation}\label{(43)}
\frac{\partial \sigma_{\mathbf{\theta}}^{2}}{\partial
Q_{1}}=-\frac{1}{B_{0}}[\frac{b}{1+a}\alpha_{1}^{2}\sigma_{\theta_{1}}^{4}+\left(\frac{b}{1+a}\right)^{2}
\alpha_{1}^{2}\alpha_{2}^{2}\sigma_{\theta_{1}}^{4}\sigma_{\theta_{1}}^{2}Q_{2}]<0,
\end{equation}
\begin{equation}\label{(44)}
\frac{\partial \sigma_{\mathbf{h}}^{2}}{\partial
Q_{1}}=-\frac{1}{B_{0}}[\frac{b}{1+a}\alpha_{2}^{2}\sigma_{\theta_{2}}^{4}+\left(\frac{b}{1+a}\right)^{2}
\alpha_{1}^{2}\alpha_{2}^{2}\sigma_{\theta_{1}}^{2}\sigma_{\theta_{1}}^{4}Q_{1}]<0,
\end{equation}
respectively. Obviously, $\sigma_{\mathbf{\theta}}^{2}$
monotonically decreases with the training power $Q_{i}$. Therefore,
the optimal power allocation requires that both desired nodes should
transmit the training sequences with their maximum allowable
transmit power.

\section{Proof of Proposition 2}

Defining $\boldsymbol{\theta}_{r}=[\mathfrak{R}\{\boldsymbol{\theta}\}^{T},
\mathfrak{I}\{\boldsymbol{\theta}\}^{T}]^{T}$, the complex Fisher
Information Matrix (FIM) can be expressed as
\begin{equation}
\mathcal{\mathbf{F}}_{uv}=\mathcal{E}\bigg\{\left(\frac{\partial
\textup{log}p(\mathbf{z}_{3}|\boldsymbol{\theta})}{u^{*}}\right)\left(\frac{\partial
\textup{log}p(\mathbf{z}_{3}|\boldsymbol{\theta})}{v^{*}}\right)^{H}\bigg\}.
\end{equation}

Following \cite{27}, the CRLB of $\boldsymbol{\theta}_{r}$ is
\begin{equation}
\textup{CRLB}=\mathcal{\mathbf{F}}_{\boldsymbol{\theta}_{r}\boldsymbol{\theta}_{r}}^{-1},
\end{equation}
where
$\mathcal{\mathbf{F}}_{\boldsymbol{\theta}_{r}\boldsymbol{\theta}_{r}}=\mathcal{\mathbf{M}}
\begin{bmatrix}\mathcal{\mathbf{F}}_{\boldsymbol{\theta}\boldsymbol{\theta}}&
\mathcal{\mathbf{F}}_{\boldsymbol{\theta}\boldsymbol{\theta}^{*}}\\
\mathcal{\mathbf{F}}_{\boldsymbol{\theta}\boldsymbol{\theta}^{*}}^{*}&\mathcal{\mathbf{F}}_{\boldsymbol{\theta}
\boldsymbol{\theta}}^{*}
\end{bmatrix}\mathcal{\mathbf{M}}^{H},\quad
\mathcal{\mathbf{M}}=\begin{bmatrix}\mathbf{I}&\mathbf{I}\\-j\mathbf{I}&j\mathbf{I}\end{bmatrix},$

The derivatives of the log-likelihood function with respect to
$\theta_{1}$ and $\theta_{2}$ can be derived as
\begin{equation}
\begin{split}
&\frac{\partial
\textup{log}p(\mathbf{z}_{3}|\boldsymbol{\theta})}{\theta_{1}^{*}}
=\alpha_{1}\alpha_{2}\tilde{\alpha}_{3}(\alpha_{1}\tilde{\alpha}_{3}h_{1}h_{2}
\tilde{\mathbf{n}}_{r}^{(1)}+\bar{\mathbf{n}})^{H}\mathbf{R}_{\tilde{\mathbf{n}}}^{-1}\mathbf{t}_{2}.
\end{split}
\end{equation}

Substituting the derivatives (74) into the FIM in (72),
$\mathcal{\mathbf{F}}_{\boldsymbol{\theta}\boldsymbol{\theta}}$ and
$\mathcal{\mathbf{F}}_{\boldsymbol{\theta}\boldsymbol{\theta}^{*}}$ can be
written in an abbreviated form:
\begin{equation}
\mathcal{\mathbf{F}}_{\boldsymbol{\theta}\boldsymbol{\theta}}=\begin{bmatrix}D_{1}&D_{2}\\D_{2}^{*}&D_{3}\end{bmatrix},\quad
\mathcal{\mathbf{F}}_{\boldsymbol{\theta}\boldsymbol{\theta}^{*}}=\begin{bmatrix}D_{4}&0\\0&0\end{bmatrix},
\end{equation}
respectively. Then $\textup{CRLB}_{\theta_{1}}$ and
$\textup{CRLB}_{\theta_{2}}$ can be derived from the CRLB by
\begin{align}
&\textup{CRLB}_{\theta_{1}}=[\textup{CRLB}]_{11}+[\textup{CRLB}]_{33},\\
&\textup{CRLB}_{\theta_{2}}=[\textup{CRLB}]_{22}+[\textup{CRLB}]_{44}.
\end{align}

Consequently, the results of \eqref{CRLB_1} and \eqref{CRLB_2} are
proven.

\section{Proof of Lemma 3}

As $D_{2}$ is only related to $|\rho|$, derivatives of
$\textup{CRLB}_{\theta_{1}}$ and $\textup{CRLB}_{\theta_{2}}$ can be
taken with respect to $|D_{2}|^{2}$ as
\begin{align}
&\frac{\partial \textup{CRLB}_{\theta_{1}}}{\partial |D_{2}|^{2}}=\frac{D_{3}\left[(|D_{2}|^{2}-D_{1}D_{2})^{2}+|D_{4}|^{2}D_{3}^{2}\right]}
{\left[(|D_{2}|^{2}-D_{1}D_{3})^{2}-|D_{4}|^{2}D_{3}^{2}\right]^{2}}>0,&\label{deta_CRLB_1}\\
&\frac{\partial \textup{CRLB}_{\theta_{2}}}{\partial
|D_{2}|^{2}}=\frac{D_{1}|D_{2}|^{4}-2D_{3}(D_{1}^{2}-|D_{4}|^{2})|D_{2}|^{2}}{\left[(|D_{2}|^{2}-D_{1}D_{3})^{2}
-|D_{4}|^{2}D_{3}^{2}\right]^{2}}&\nonumber\\
&+\frac{D_{1}^{2}D_{3}^{2}(D_{1}^{2}-|D_{4}|^{2})}{\left[(|D_{2}|^{2}-D_{1}D_{3})^{2}-|D_{4}|^{2}D_{3}^{2}\right]^{2}}>0.&\label{deta_CRLB_2}
\end{align}

Obviously, $\textup{CRLB}_{\theta_{1}}$ is a increasing function of $|\rho|^{2}$
due to the fact that \eqref{deta_CRLB_1} is positive. When $|\rho|=0$, $\textup{CRLB}_{\theta_{1}}$ reaches its
minimum value. While for $\textup{CRLB}_{\theta_{2}}$, its
numerator can be re-expressed as
\begin{equation}
D_{1}\left[|D_{2}|^{2}-\frac{D_{3}}{D_{1}}(D_{1}^{2}-|D_{4}|^{2})\right]^{2}+\frac{D_{3}^{2}}{D_{1}}(D_{1}^{2}-|D_{4}|^{2})|D_{4}|^{2}.
\end{equation}

When $|D_{4}|^{2}<D_{1}^{2}$, \eqref{deta_CRLB_2} is positive. Therefore,
both $\textup{CRLB}_{\theta_{1}}$ and $
\textup{CRLB}_{\theta_{2}}$ are increasing functions of
$|\rho|^{2}$, which achieve their minimum values at
$|\rho|=0$. Namely, the optimal $\mathbf{t}_{1}$ and
$\mathbf{t}_{2}$ should be orthogonal. When $|\rho|=0$, the
corresponding CRLBs can be expressed as
\begin{align}
&\textup{CRLB}_{\theta_{1}}=\frac{D_{1}}{D_{1}^{2}-|D_{4}|^{2}},\\
&\textup{CRLB}_{\theta_{2}}=\frac{1}{D_{3}}.
\end{align}

It is supposed that the maximum power assigned to $\mathbf{t}_{i}$
is $Q_{i}^{max}$ (typically $LP_{i}$). By minimizing CRLBs, the
optimal training power assigned for each $\mathbf{t}_{i}$ can be
achieved. The optimal training power for
$\textup{CRLB}_{\theta_{1}}$ is $LP_{r2}$, which means the desired
node $\mathbb{T}_{2}$ needs to transmit with its maximum allowable
transmit power. For $\textup{CRLB}_{\theta_{2}}$, the corresponding
optimization for $Q_{1}$ is
\begin{equation}
Q_{1}=\arg \min_{Q_{1}} \frac{Q_{1}+c}{Q_{1}^{2}+2cQ_{1}},
\end{equation}
where $c=\frac{a_{0}\sigma_{n}^{2}(r-2)^{2}}{4(1+a)}$. Since the
objective function is a decreasing function of $Q_{1}$, the optimal
training power for $\mathbf{T}_{1}$ is achieved with the maximum
allowable transmit power $Q_{1}^{max}$.

\section{Proof of Proposition 4}

Firstly, the corresponding covariance matrix of the equivalent noise
is
\begin{equation}
\begin{split}
&\mathbf{R}_{\tilde{\mathbf{n}}_{z}}=\bigg\{1+\sideset{}{_{i=1}^{N-1}}\sum
\left(\sideset{}{_{j=i}^{N-1}}\prod\alpha_{2j-1}^{2}\sigma_{2j-1}^{2}\right)\\
&+\sideset{}{_{i=1}^{N-1}}\sum\big[\sideset{}{_{k=1}^{N}}\prod\alpha_{2k-1}^{2}
\sigma_{2k-1}^{2}\left(\sideset{}{_{j=i}^{N-1}}\prod\alpha_{2j}^{2}\sigma_{2j}^{2}\right)+\alpha_{2N-1}^{2}\sigma_{1}^{2}\\
&\sideset{}{_{k=1}^{i-1}}\prod\alpha_{2k-3}^{2}\sigma_{2k-1}^{2}\left(2^{N-i}
\sideset{}{_{j=i}^{N-1}}\prod\alpha_{2j-1}^{4}\sigma_{2j+1}^{4}\right)
\big]\bigg\}\sigma_{n}^{2}\mathbf{I}.
\end{split}
\end{equation}

As the number of nodes $N$ increases,
$\mathbf{R}_{\tilde{\mathbf{n}}_{z}}$ reaches the upper bound under
certain circumstances. The transmit power is assumed to be $P$, and
the channel gains are assumed to have the same variance $\sigma$. On
denoting $\alpha_{i}^{2}\sigma_{i}^{2}$ by $\omega$, and
substituting the simplified expression into
$\mathbf{R}_{\tilde{\mathbf{n}}_{z}}$, (84) can be rewritten as
\begin{equation}
\mathbf{R}_{\tilde{\mathbf{n}}_{z}}=\big[\frac{1-\omega^{N}+\omega^{N+1}-\omega^{2N}}
{1-\omega}+\frac{2\omega^{N+1}\left[1-(2\omega)^{N-1}\right]}{1-2\omega}\big]\sigma_{n}^{2}\mathbf{I}.
\end{equation}

When $N$ approaches to infinity, the upper bound of
$\mathbf{R}_{\tilde{\mathbf{n}}_{z}}$ can be written as
\begin{equation}
\lim_{n\rightarrow \infty}
\mathbf{R}_{\tilde{\mathbf{n}}_{z}}=\frac{1}{1-\omega}\sigma_{n}^{2}\mathbf{I}.
\end{equation}

Similarly to the $4-$hop WCNs, the performance of LMMSE for the
$2N-$hop case can be derived conditionally. The MSE of
$\mathbf{h}_{\mathbf{z}}$ can be expressed as
\begin{align}
\sigma_{\mathbf{h}_{\mathbf{z}}}^{2}&=\textup{tr}\bigg\{\left(\mathbf{R}_{\mathbf{h}_{\mathbf{z}}}^{-1}
+\sideset{}{_{i=1}^{N}}\prod\alpha_{2i-1}\mathbf{\Lambda}_{\mathbf{z}}\mathbf{T}^{H}\mathbf{R}
_{\tilde{\mathbf{n}}_{z}}^{-1}\mathbf{T}\mathbf{\Lambda}_{\mathbf{z}}\right)^{-1}\bigg\}\nonumber\\
&=\textup{tr}\bigg\{\left(\textup{diag}\{2^{N}\sigma^{4N},
\sigma^{4N}\}^{-1}+\frac{\kappa^{2N-1}}{\eta\sigma_{n}^{2}}\mathbf{T}^{H}\mathbf{T}\right)^{-1}\bigg\},
\end{align}
where $\kappa$ represents $\alpha_{i}^{2}$. If $\kappa<1$, the
parameter $\frac{\kappa^{2n-1}}{\eta\sigma_{n}^{2}}$ is equal to
zero, which indicates that the MSE can be expressed as
$\sigma_{\mathbf{h}_{\mathbf{z}}}^{2}=(1+2^n)\sigma^{4n}$. If
$\sigma^{2}>\sqrt{2}/2$, there is no upper bound for the MSE. When
$\sigma^{2}=\sqrt{2}/2$, the MSE is simply equal to one. Obviously,
the MSE approaches zero if the variance satisfies
$\sigma^{2}<\sqrt{2}/2$. If the amplification factor is equal to
one,
$\frac{\kappa^{2n-1}}{\eta\sigma_{n}^{2}}=\frac{1-\omega}{\sigma_{n}^{2}}$
remains constant. When $\sigma^{2}>1$, the MSE performance of
$\varpi_{1}$ and $\varpi_{2}$ can be derived as
$\sigma_{\varpi_{1}}^{2}=\frac{\sigma_{n}^{2}}{(1-\omega)(1-\rho^{2})Q_{1}}$ and
$\sigma_{\varpi_{2}}^{2}=\frac{\sigma_{n}^{2}}{(1-\omega)(1-\rho^{2})Q_{2}}$,
respectively.

In the case of the amplification factor $\kappa>1$, the parameter
$\frac{\kappa^{2n-1}}{\eta\sigma_{n}^{2}}$ approaches infinity
and the MSE cannot be obtained directly. Similarly, we can derive the
CRLBs for $2N$-hop WCNs in the same way we did for $4-$hop WCNs. In
brief, only the case when the amplification $\alpha_{i}=1$ is
considered; the coefficients $D_{i}$ in the CRLB can be transformed as $
D_{1}=\frac{(1-\omega)Q_{1}}{\sigma_{n}^{2}}$,
$D_{2}=\frac{\rho(1-\omega)\sqrt{Q_{1}Q_{2}}}{\sigma_{n}^{2}}$,
$D_{3}=\frac{(1-\omega)Q_{2}}{\sigma_{n}^{2}}$, and $\quad D_{4}=0$.

The CRLB for $\varpi_{1}$ and $\varpi_{2}$ in $2N-$hop WCNs can be
derived respectively as
$\textup{CRLB}_{\varpi_{1}}\!=\!\frac{\sigma_{n}^{2}}{(1-\omega)(1-\rho^{2})Q_{1}}$ and
$\textup{CRLB}_{\varpi_{2}}\!=\!\frac{\sigma_{n}^{2}}{(1-\omega)(1-\rho^{2})Q_{2}}$.
It is noted that CRLBs are almost the same as the MSE. Moreover,
the same conclusions in $4-$hop WCNs can be extended to $2N-$hop
WCNs in special cases when the optimal training is orthogonal and
the maximum allowable transmit power is used.

\begin{IEEEbiography}{Mugen Peng}
(M'05--SM'11) received the Ph.D. degree in communication and
information system from the Beijing University of Posts \&
Telecommunications (BUPT), China, in 2005. After the Ph.D.
graduation, he joined BUPT, and became a full professor with the
School of Information and Communication Engineering, BUPT, in
October 2012. During 2014, he is also an academic visiting fellow
with Princeton University, Princeton, NJ, USA. He is leading a
research group focusing on wireless transmission and networking
technologies in the Key Laboratory of Universal Wireless
Communications (Ministry of Education) at BUPT. His main research
areas include wireless communication theory, radio signal processing
and convex optimizations, with particular interests in cooperative
communication, radio network coding, self-organizing network,
heterogeneous network, and cloud computing. He has
authored/coauthored more than 40 refereed IEEE journal papers and
over 200 conference proceeding papers.

Dr. Peng is currently on the Editorial/Associate Editorial Board of
\emph{IEEE Communications Magazine}, \emph{IEEE Access},
\emph{International Journal of Antennas and Propagation} (IJAP),
\emph{China Communication}, and \emph{International Journal of
Communications System }(IJCS). He has been the guest leading editor
for special issues of \emph{IEEE Wireless Communications}, IJAP
and the \emph{International Journal of Distributed Sensor Networks}
(IJDSN). Dr. Peng was honored with the 2014 IEEE ComSoc AP
Outstanding Young Researcher Award, and the Best Paper Award in
GameNets 2014, CIT 2014, ICCTA 2011, IC-BNMT 2010, and IET CCWMC
2009. He was awarded the First Grade Award of Technological
Invention by the Ministry of Education of China for his excellent
research work on hierarchical cooperative communication theory
and technologies, and the Second Grade Award of Scientific \&
Technical Progress from the China Institute of Communications for his
excellent research work on the co-existence of multi-radio access
networks and 3G spectrum management in China.
\end{IEEEbiography}

\begin{IEEEbiography}{Qiang Hu}
received the B.S. degree in applied physics from the Beijing University
of Posts \text{\&} Communications (BUPT), China, in 2013. He is
currently pursuing the Master degree at BUPT. His research interests
include cooperative communications, such as cloud radio access
networks (C-RANs), and statistical signal processing in large-scale
networks.
\end{IEEEbiography}

\begin{IEEEbiography}{Xinqian Xie}
received the B.S. degree in telecommunication engineering from
the Beijing University of Posts and Communications (BUPT), China, in
2010. He is currently pursuing the Ph.D. degree at BUPT. His
research interests include cooperative communications, estimation
and detection theory.
\end{IEEEbiography}

\begin{IEEEbiography}{Zhongyuan Zhao} is currently a lecturer with the Key Laboratory of
Universal Wireless Communication (Ministry of Education) at the Beijing
University of Posts \& Telecommunications (BUPT), China. He received
his Ph.D. degree in communication and information systems and B.S.
degree in applied mathematics from BUPT in 2009 and 2014,
respectively. His research interests include network coding, MIMO,
relay transmissions, and large-scale cooperation in future
communication networks.
\end{IEEEbiography}

\begin{IEEEbiography}{H. Vincent Poor}
(S'72, M'77, SM'82, F'87) received the Ph.D. degree in EECS from
Princeton University in 1977.  From 1977 until 1990, he was on the
faculty of the University of Illinois at Urbana-Champaign. Since
1990 he has been on the faculty at Princeton, where he is the
Michael Henry Strater University Professor of Electrical Engineering
and Dean of the School of Engineering and Applied Science. Dr.
Poor's research interests are in the areas of stochastic analysis,
statistical signal processing, and information theory, and their
applications in wireless networks and related fields such as social
networks and smart grid. Among his publications in these areas are
the recent books \textit{Principles of Cognitive Radio} (Cambridge
University Press, 2013) and \textit{Mechanisms and Games for Dynamic
Spectrum Allocation} (Cambridge University Press, 2014).

Dr. Poor is a member of the National Academy of Engineering and the
National Academy of Sciences, and a foreign member of Academia
Europaea and the Royal Society. He is also a fellow of the American
Academy of Arts and Sciences, the Royal Academy of Engineering
(U.K.), and the Royal Society of Edinburgh. He received the
Marconi and Armstrong Awards of the IEEE Communications Society in 2007 and 2009,
respectively. Recent
recognition of his work includes the 2014 URSI Booker Gold Medal,
and honorary doctorates from several universities in Europe and
Asia, including an honorary D.Sc. from Aalto University in 2014.
\end{IEEEbiography}


\begin{thebibliography}{8}
\bibitem{1}
N. Kayastha, D. Niyato, P. Wang, and E. Hossain,\textquotedblleft
Applications, architectures, and protocol design issues for mobile
social networks: A survey,\textquotedblright~\emph{Proceedings of
IEEE}, vol. 99, no. 12, pp. 2130--2158, Dec. 2011.

\bibitem{2}
A. Weaver, and B. Morrison, \textquotedblleft Social
networking,\textquotedblright~\emph{IEEE Computer}, vol. 41, no. 2,
pp. 97--100, Feb. 2008.

\bibitem{7}
E. Stai, V. Karyotis, and S. Papavassiliou, \textquotedblleft
Topology enhancements in wireless multi-hop networks: A top-down
approach,\textquotedblleft~\emph{IEEE Trans. Parallel and
Distributed Systems}, vol. 23, no. 7, pp. 1344--1357, July 2012.

\bibitem{8}
K. C. Chen, M. Chiang, and H. V. Poor, \textquotedblleft From
technological networks to social
networks,\textquotedblright~\emph{IEEE J. Sel. Areas Commun.}, vol.
31, no. 9, pp. 548¨C-572, Sep. 2013.

\bibitem{9}
A. Eryilmaz and R. Srikant, \textquotedblleft Joint congestion
control, routing and MAC for stability and fairness in wireless
networks,\textquotedblright~\emph{IEEE J. Sel. Areas Commun.}, vol.
24, no. 8, pp. 1514--1524, Aug. 2006.

\bibitem{10}
K. C. Chen, H. V. Poor, and R. Prasad, \textquotedblleft Mobile social
networks [Guest Editorial],\textquotedblright~\emph{IEEE Wireless
Commun.}, vol. 21, no. 1, pp. 8--9, Feb. 2014.

\bibitem{11}
X. Liang; K. Zhang, X. Shen, and X. Lin, \textquotedblleft Security
and privacy in mobile social networks: challenges and
solutions,\textquotedblright~\emph{IEEE Wireless Commun.}, vol. 21,
no. 1, pp. 33--41, Feb. 2014.

\bibitem{12}
J. Munoz, J. Garzon, P. Ameigeiras, J. Ortiz, and J. Soler,
\textquotedblleft Characteristics of mobile youtube
traffic,\textquotedblright~\emph{IEEE Wireless Commun.}, vol. 21,
no. 1, pp. 18--25, Feb. 2014.

\bibitem{13}
E. Stai, V. Karyotis, and S. Papavassiliou, \textquotedblleft
Exploiting socio-physical network interactions via a utility-based
framework for resource management in mobile social
networks,\textquotedblright~\emph{IEEE Wireless Commun.}, vol. 21,
no. 1, pp. 10--17, Feb. 2014.
\bibitem{14}
A. Sendonaris, E. Erkip,  and B. Aazhang, \textquotedblleft User
cooperation diversity: Part I. System
description,\textquotedblright~\emph{IEEE Trans. Commun.}, vol. 51,
no. 11, pp. 1927--1938, Nov. 2003.
\bibitem{15}
A. Nosratinia, T. Hunter, and A. Hedayat, \textquotedblleft
Cooperative communication in wireless networks,\textquotedblright~
\emph{IEEE Commun. Magazine}, vol. 42, no. 10, pp. 74--80, Oct.
2004.
\bibitem{16}
R. Ahlswede, N. Cai, S. Li, and R. Yeung, \textquotedblleft Network
information flow,\textquotedblright~\emph{IEEE Trans. Inform.
Theory}, vol. 46, no. 4, pp. 1204--1216, July 2000.
\bibitem{17}
S. Sengupta, S. Rayanchu, and S. Banerjee, \textquotedblleft Network
coding-aware routing in wireless
networks,\textquotedblright~\emph{IEEE/ACM Trans. Networking}, Vol.
18, No. 4, pp. 1158-1170, Aug. 2010.
\bibitem{18}
R. Prior, D. Lucani, Y. Phulpin, M. Nistor, and J. Barros,
\textquotedblleft Network coding protocols for smart grid
communications,\textquotedblright~\emph{IEEE Trans. Smart Grid},
2014.
\bibitem{19}
N. S. Riberio J\'{u}nior, \emph{et al.}, \textquotedblleft CodeDrip:
Data dissemination protocol with network coding for wireless sensor
networks,\textquotedblright~\emph{Wireless Sensor Networks},
Springer International Publishing, pp: 34-49, 2014.
\bibitem{20}
P. Pahlavani, D. ELucani, M. Pedersen, and F. Fitzek.
\textquotedblleft PlayNCool: opportunistic network coding for local
Optimization of routing in wireless mesh
networks,\textquotedblright~in \emph{Proc. IEEE GLOBECOM Workshops}, Alatanta, GA, Dec. 2013,
pp. 812--817.
\bibitem{21}
F. Fitzek, J. Heide, M. V. Pedersen, and M. Katz, \textquotedblleft
Implementation of network coding for social mobile
clouds,\textquotedblright~\emph{IEEE Signal Processing Magazine}
vol. 30, no. 1, Jan. 2013.
\bibitem{22}
S. Li, R. Yeung, and N. Cai, \textquotedblleft Linear network
coding,\textquotedblright~\emph{IEEE Trans. Inform. Theory}, vol.
49, no. 2, pp. 371--381, Feb. 2003.
\bibitem{23}
F. Gao, R. Zhang, and Y. Liang, \textquotedblleft Optimal channel
estimation and training desing for two-way relay
networs,\textquotedblright~\emph{IEEE Trans. Commun.}, vol. 57, no.
10, pp. 3024-3033, Oct. 2009.
\bibitem{24}
B. Jiang, F. Gao, X. Gao, and A. Nallanathan, \textquotedblleft
Channel estimation and training desing for two-way relay networks
with power allocation,\textquotedblright~\emph{IEEE Trans. Wireless
Commun.}, vol. 9, no. 6, pp. 2022--2032, Jun. 2010.
\bibitem{25}
X. Xie, M. Peng, B. Zhao, W. Wang, and Y. Hua, \textquotedblleft
Maximum a posteriori based channel estimation strategy on two-way
relaying channels,\textquotedblright~\emph{IEEE Trans. Wireless
Communi.}, vol. 13, no. 1, pp. 450--463, Jan. 2014.
\bibitem{WSN_Hist1}
F. Tabataba, P. Sadeghi, C. Hucher, and M. Pakravan,
\textquotedblleft Impact of channel estimation errors and power
allocation on analog network coding and routing in two-way
relaying,\textquotedblright~\emph{IEEE Trans. Vehicular Tech.}, vol.
61, no. 7, pp. 3223--3239, Sep. 2012.
\bibitem{26}
S. Kay, \emph{Fundamentals of Statistical Signal
Processing: Estimation Theory}, Englewood
Cliffs NJ: Prentice Hall, 1987.

\bibitem{28}
M. Sichitiu and C. Veerarittiphan, \textquotedblleft Simple,
accurate time synchronization for wireless sensor
networks,\textquotedblright~in \emph{Proc. IEEE Wireless Communications and
Networking Conference (WCNC)}, New Orleans, LA, Mar. 2003, pp. 1266--1273.

\bibitem{27}
E. de Carvalho, J. Cioffi, and D. Slock, \textquotedblleft
Cram\'{e}r-rao bounds for blind multichannel
estimation,\textquotedblright~in \emph{Proc. IEEE GLOBECOM}, San
Francisco, CA, Nov. 2000, pp. 1036--1040.
\end{thebibliography}
\end{document}